\documentclass[10pt, twocolumn, comsoc]{IEEEtran}

\usepackage{graphicx,epsfig}
\usepackage[noadjust]{cite}
\usepackage{mcite}
\usepackage{amsfonts,helvet}
\usepackage{fancyhdr}
\usepackage{threeparttable}
\usepackage{epsf,epsfig}
\usepackage{amsthm}
\usepackage{amsmath}
\usepackage{siunitx}
\usepackage{amssymb}
\usepackage{stfloats}

\usepackage[colorlinks=true, linkcolor=blue]{hyperref}

\usepackage{dsfont}
\usepackage{subfig}
\usepackage{color}
\usepackage{enumerate}
\usepackage{gensymb}
\usepackage{cancel}
\usepackage{lipsum}
\usepackage{mathtools}
\usepackage{cuted}
\usepackage{bbm}
\usepackage[linesnumbered,ruled]{algorithm2e}

\newtheorem{lemma}{Lemma}

\newtheorem{remark}{Remark}

\usepackage{eucal}

\newcommand{\herm}{{\sf H}}
\newcommand{\trans}{{\sf T}}

\def\bbC{\mbox{$\mathbb{C}$}}

\newcommand{\bfR}{{\mathbf{R}}}
\newcommand{\bfC}{{\mathbf{C}}}
\newcommand{\bfL}{{\mathbf{L}}}

\newcommand{\bfX}{{\mathbf{X}}}

\newcommand{\bfU}{{\mathbf{U}}}

\newcommand{\bfV}{{\mathbf{V}}}
\newcommand{\bfv}{{\mathbf{v}}}

\begin{document}

\title{Fractional Programming for Kullback-Leibler Divergence in Hypothesis Testing}

\author{Jeongwoo~Park, Seongkyu~Jung, Kaiming~Shen, and Jeonghun~Park

\thanks{This work was supported in part by Institute of Information \& Communications Technology Planning \& Evaluation (IITP) grant funded by the Korea government (MSIT) (No. RS-2024-00397216, Development of the Upper-mid Band Extreme massive MIMO (E-MIMO), No. 2025-RS-2024-00428780, 6G Cloud Research and Education Open Hub and No. RS-2024-00434743, YKCS Open RAN Global Collaboration Center).
J. Park, S. Jung and J. Park are with School of Electrical and Electronic Engineering, Yonsei University, Seoul, South Korea (e-mail:{\texttt{\{simyffff, wjdtjd963, jhpark\}@yonsei.ac.kr}}). K. Shen is with School of Science and Engineering, The Chinese University of Hong Kong, Shenzhen, China (e-mail: {\texttt{shenkaiming@cuhk.edu.cn}}).
}
}

\maketitle \setcounter{page}{1} 
\begin{abstract} 
Maximizing the Kullback-Leibler divergence (KLD) is a fundamental problem in waveform design for active sensing and hypothesis testing, as it directly governs to the error exponent of detection probability. However, the associated optimization problem is highly nonconvex due to the intricate coupling of log-determinant and matrix trace terms. Existing solutions often suffer from prohibitively high computational complexity, typically requiring matrix inversion at every iteration. In this paper, we propose a computationally efficient optimization framework based on fractional programming (FP). Our key idea is to reformulate the KLD maximization problem as a sequence of tractable quadratic subproblems using matrix FP. To further reduce complexity, we introduce a nonhomogeneous relaxation technique that replaces the costly linear system solver with a simple closed-form update, thereby reducing the per-iteration complexity to quadratic order. To compensate for the convergence speed trade-off induced by relaxation, we employ an acceleration method called STEM by interpreting the iterative scheme as a fixed-point mapping. The resulting algorithm achieves significantly faster convergence rates with low per-iteration cost. Numerical results demonstrate that our approach reduces the total runtime by orders of magnitude compared to a state-of-the-art benchmark. Finally, we apply the proposed framework to a multiple random access scenario and a joint integrated sensing and communication scenario, validating the efficacy of our framework in such applications. 

\end{abstract}

\section{Introduction}

Kullback--Leibler divergence (KLD), also known as relative entropy, quantifies the amount of dissimilarity between two probability distributions \cite{KLD51}. 
As a measure of how one distribution diverges from another in expectation, 
KLD plays a pivotal role in various formulations of information theory. 
For example, in mismatched source coding, a KLD-related quantity, i.e., the cross-entropy, naturally characterizes the rate penalty incurred when encoding is performed under an incorrect source model \cite{merhavTIT94}. 
In universal compression, KLD determines the redundancy rate when the encoder must operate without precise knowledge of the source distribution. 
In particular, considering hypothesis testing, which is the main focus of this paper, maximizing the KLD between the hypotheses directly enhances their statistical separability. 
According to the Chernoff–Stein lemma \cite{chernoff}, the KLD dictates the error exponent of the miss-detection probability under a fixed false-alarm constraint; hence, a larger KLD implies a faster exponential decay of detection errors. 
From this perspective, maximizing the KLD leads to sensing waveforms that maximize detection probability under given long-term statistics. 
Such designs directly determine the sensing system’s ability to reliably detect targets 
under practical resource constraints, e.g., sensing time, bandwidth, or power. 
Accordingly, the development of efficient KLD optimization techniques is of central importance in sensing systems.

The recent rise of integrated sensing and communication (ISAC) systems further intensifies the need for such efficient KLD design methods. 
Given that sensing and communication share spectrum, power, and front-end hardware in ISAC, 
the transmit waveforms must be jointly designed to simultaneously optimize communication and sensing performance metrics.
% mutual information and statistical sensing separability. 
In this context, KLD is often employed as a sensing performance metric, while mutual information (MI) is typically used to characterize the achievable rates \cite{al2023unified, fei2024revealing, kloob2024novel}. 
Such a joint design objective demands tractable and scalable optimization algorithms. 

However, the analytical form of the KLD is often highly intricate, making direct optimization intractable. 
Even under the Gaussian assumption, which yields an explicit closed-form expression, the resulting optimization problem typically remains nonconvex due to the intricate coupling between the waveform parameters and the statistical structure of the received signal. 
This nonconvexity limits the applicability of conventional convex optimization techniques since neither global optimality nor reliable convergence properties can be guaranteed. 
In this paper, we propose a new optimization method for maximizing the KLD in hypothesis testing. Our key idea is to apply fractional programming (FP) \cite{shenFPtsp1, shenFPtsp2} to reformulate the problem into a sequence of concave subproblems, whereby each subproblem can be solved efficiently. We demonstrate that our method achieves superior performance compared to existing state-of-the-art methods, while significantly reducing computational complexity.

\subsection{Related Works}

Prior works addressed the sensing waveform design problem primarily in the context of multi-input multi-output (MIMO) radar systems. 
Under this setup, existing studies predominantly focused on optimizing estimation-theoretic criteria or signal-to-interference-plus-noise ratio (SINR). 
For instance, \cite{stoica2007probing} suggested shaping the probing signal covariance matrix to approximate a desired beampattern or to maximize the spatial power at target locations, thereby enhancing the SINR. 
In \cite{bekkerman2006target}, it was demonstrated that spatially orthogonal signal transmission minimizes the Cram\'{e}r-Rao bound (CRB) for direction-of-arrival estimation. These works primarily aim at sensing tasks formulated as parameter estimation problems. 
In \cite{bell:tit:93}, it was shown that maximizing the MI between the random target response and the received signal is equivalent to minimizing the mean-squared error (MSE) of the target response estimation.

Another related line of research concerns nonconvex low-autocorrelation sequence design for radar and communication systems. In this literature, the goal is typically to suppress autocorrelation sidelobes by minimizing integrated sidelobe level (ISL), weighted integrated sidelobe level (WISL), peak sidelobe level (PSL), or related metrics under practical sequence constraints. For example, majorization-minimization-based frameworks have been developed for unified ISL/WISL/PSL-type objectives under unimodular, peak-to-average-power ratio, similarity, and discrete-phase constraints, with efficient updates based on fast Fourier transform/inverse fast Fourier transform operations~\cite{zhao2016unified}. Power method-like iteration (PMLI)-based approaches have also been proposed to solve WISL minimization by transforming the quartic autocorrelation objective into quadratic or bi-quadratic subproblems~\cite{eamaz2023cypmli}. More recently, relaxation-based PMLI methods have been developed for finite-alphabet or discrete-phase WISL minimization~\cite{eamaz2023marli}. These works provide useful methodological context for the present study, particularly in their use of surrogate-based and fixed-point-type optimization ideas, while the KLD-oriented objective and matrix waveform design setting considered here lead to a different problem structure.

In contrast to these estimation- and sidelobe-oriented design criteria, distinguishing between hypotheses (i.e., detection) is fundamentally different from parameter estimation. For this reason, when the primary objective is detection, the adoption of KLD as a performance metric is rigorously motivated.
The use of KLD for detection-centric waveform design originates from \cite{zhu2017information}. Therein, the locally most powerful (LMP) detector was derived under low signal-to-noise ratio (SNR) conditions, where the LMP detector is known to be optimal. 
It was shown that the detection performance of the LMP detector is strictly monotonic in the KLD, thereby establishing the KLD as a rigorous optimality criterion for target detection.

Nonetheless, practical optimization algorithms for KLD are much less sophisticated compared to those for other information-theoretic metrics, such as MI \cite{shenFPtsp1, shen2019optimization, parkGPIRS23, christensen2008weighted, shi2011iteratively}. 
While MI maximization has been extensively studied and often admits elegant closed-form solutions via water-filling algorithms \cite{Yang:taes:07}, the analytical form of KLD for detection involves intricate log-determinant (log-det) and trace terms coupled with the waveform covariance, rendering KLD maximization highly nonconvex and intractable. To circumvent this intractability, earlier approaches relied on restrictive assumptions or heuristic designs. For instance, in \cite{wang2013adaptive}, the authors considered optimizing parameters for the fixed waveform structure (e.g., chirp rate or pulse width) to improve resolvability, rather than optimizing the waveform structure itself. Similarly, \cite{grossi2012space} derived analytical solutions for space-time codes, but their validity was limited to specific spectral conditions or global energy constraints.

Efforts to address general nonconvex KLD problems without such restrictions have focused on numerical optimization techniques. \cite{wang2018multi} modeled the problem as sequential multi-hypothesis testing and employed semidefinite relaxation (SDR) to handle the nonconvex constraints. While effective, SDR-based methods suffer from high computational complexity and require randomization steps to recover rank-1 solutions. 
The current state-of-the-art is represented by \cite{tang:tsp:18}, which applied the minorization-maximization (MM) framework. In the sequel, we refer to the method as the adaptive waveform-design MM algorithm, abbreviated as AWD-MM. Although its quadratic minorizer reduces the per-iteration complexity compared to previous approaches \cite{tang2015relative} proposed by the same authors, it still necessitates solving a large-scale linear system at each iteration, which imposes a cubic computational burden that restricts scalability for large antenna arrays.

Beyond these KLD-specific solvers, broader matrix-ratio optimization frameworks---particularly FP---have proven highly effective for communication and sensing-related problems. Scalar, vector and matrix FP techniques have been used to handle SINR, rate, weighted-sum-rate, beamforming, scheduling, and related matrix-ratio objectives in wireless communication systems \cite{shenFPtsp1, shenFPtsp2, shen2019optimization}. Recent ISAC studies have also exploited FP-type reformulations for communication-rate and Fisher-information based beamforming, together with nonhomogeneous bounds to reduce the cost of large matrix inversions \cite{chen2025fast}. However, the Gaussian KLD objective considered in this paper is not a standard MI, SINR-ratio, or Fisher-information expression. It exhibits a more intricate coupling between the waveform and multiple matrix-valued terms. Therefore, the Gaussian KLD structure calls for a tailored reformulation that goes beyond direct application of standard FP or MM templates.

In ISAC systems, KLD has recently been adopted as a unified performance metric for analyzing sensing-communication trade-offs \cite{al2023unified, fei2024revealing, kloob2024novel}. These studies demonstrate the relevance of KLD for ISAC, but they do not provide a scalable transmit-waveform optimization framework that directly maximizes the Gaussian KLD sensing objective jointly with a communication metric. In contrast, many existing ISAC waveform designs rely on CRB, beampattern MSE, radar SINR, Fisher-information, or MI-type sensing objectives \cite{liuFanCRLBopt22, choiGPIisac24, kim:twc:25}. Hence, there remains a gap between detection-theoretic KLD-based sensing criteria and computationally efficient waveform optimization for both pure sensing and ISAC settings. To bridge this gap, this paper develops an FP-based reformulation that converts the Gaussian KLD objective into tractable surrogate subproblems and naturally integrates with other compatible objectives such as MI, enabling joint ISAC waveform design.

\subsection{Contributions}

We propose a novel optimization framework to address KLD maximization in binary hypothesis testing, which is the primary focus of this paper.
Our key idea is to reformulate the intricate Gaussian KLD objective using FP \cite{shen2019optimization}, thereby enabling the development of an algorithm that guarantees monotonic convergence with significantly reduced computational complexity. The main contributions are summarized as follows:

\begin{itemize}
    \item We reformulate the nonconvex Gaussian KLD objective by applying matrix FP techniques to the log-det term and the trace term. This approach decouples the matrix inverse from the design variables, transforming the original problem into a sequence of concave quadratic subproblems. This reformulation exploits the structure of KLD to provide a tight surrogate function that guarantees monotonic convergence.
    \item To overcome the cubic complexity of solving linear systems in the FP approach, we introduce a nonhomogeneous relaxation technique. By replacing the anisotropic curvature of the surrogate with a conservative isotropic bound, we derive a closed-form waveform update with quadratic complexity.
    This significantly enhances scalability while strictly preserving the monotonic ascent property.
    \item We develop an accelerated optimization framework, termed A-MM-KLD, by interpreting the iterative algorithm as a fixed-point mapping and incorporating a Steffensen-type acceleration. In this scheme, the accelerator uses a secant-type approximation of the fixed-point mapping to improve the effective local convergence factor of the baseline MM iteration. This synergy effectively compensates for the increased iteration count caused by the relaxed surrogate, resulting in a substantial reduction in total runtime compared to a state-of-the-art baseline.    
    \item We establish that the proposed algorithms are valid instantiations of the MM principle. Specifically, they produce a monotonically non-decreasing sequence of objective values, and every accumulation point of the iterates is a stationary point of the original nonconvex KLD problem; global optimality, however, cannot be claimed, since the problem does not fall into the class of nonconvex problems with provably benign landscapes \cite{tu2016low, bhojanapalli2016global, candes2015phase, zhao2015nonconvex}. The unaccelerated updates have the standard local fixed-point behavior, typically linear near an attracting stationary point, while A-MM-KLD uses a safeguarded STEM step to improve the practical local convergence speed without changing the stationary-point convergence guarantee.

    \item To illustrate the generality of the proposed framework, we demonstrate that our framework is applicable to several practical scenarios, including ISAC and multiple random access scenarios. In the ISAC case, leveraging the fact that MI maximization via FP is well established \cite{shenFPtsp1, chen2025fast}, we show that our proposed framework can be seamlessly integrated to jointly maximize the composite objective of MI and KLD. In the multiple random access scenario, formulated as maximizing a weighted sum of Gaussian KLDs, each term in the objective preserves the single-Gaussian algebraic form, ensuring that our method can be suitably applied.     
\end{itemize}

\subsection{Notations}
% Boldface lower- and upper-case letters denote column vectors and matrices, respectively. $\mathbb{C}^m$ and $\mathbb{C}^{m\times n}$ denote the sets of $m$-dimensional complex vectors and $m\times n$ complex matrices, while $\mathbb{H}_+$ and $\mathbb{H}_{++}$ denote the sets of Hermitian positive semi-definite (PSD) and positive definite (PD) matrices, with the superscript $n$, when specified, indicating the dimension $n\times n$. For a matrix $\mathbf{A}$, $\mathbf{A}^{\sf T}$, $\mathbf{A}^{\sf H}$, $|\mathbf{A}|$, $\operatorname{rank}(\mathbf{A})$, and $\operatorname{span}(\mathbf{A})$ denote its transpose, Hermitian transpose, determinant, rank, and column space, whereas $\|\mathbf{a}\|$ and $\|\mathbf{A}\|_F$ denote the Euclidean and Frobenius norms. $\mathbf{I}_n$ is the $n\times n$ identity matrix and $\mathbf{0}$ the all-zero vector or matrix, with subscripts omitted when the dimension is clear from context. The operators $\operatorname{vec}(\cdot)$, $\otimes$, $\langle\cdot,\cdot\rangle$, and $\Re\{\cdot\}$ denote vectorization, the Kronecker product, the inner product, and the real part, respectively. For matrix inequalities, $\mathbf{A}\succeq\mathbf{B}$ (or $\mathbf{A}\succ\mathbf{B}$) means $\mathbf{A}-\mathbf{B}$ is PSD (or PD). Finally, $\mathbb{E}_\mathbf{X}[\cdot]$ denotes expectation over the random variables $\mathbf{X}$, and $\mathcal{CN}(\mathbf{0},\mathbf{R})$ the zero-mean circularly symmetric complex Gaussian distribution with covariance $\mathbf{R}$ and probability density function (PDF) $p_{\mathcal{CN}}(\mathbf{x};\mathbf{R})$.
Boldface lower-case and upper-case letters denote column vectors and matrices, respectively. The set of $m$-dimensional complex vectors and $m\times n$ complex matrices are denoted by $\mathbb{C}^m$ and $\mathbb{C}^{m\times n}$, while $\mathbb{H}_+$ and $\mathbb{H}_{++}$ represent the sets of Hermitian positive semidefinite (PSD) and positive definite (PD) matrices, respectively, where the superscript $n$, if specified, denotes the matrix dimension $n\times n$. For a matrix $\mathbf{A}$, $\mathbf{A}^{\sf T}$, $\mathbf{A}^{\sf H}$, $|\mathbf{A}|$, $\operatorname{rank}(\mathbf{A})$, and $\operatorname{span}(\mathbf{A})$ denote its transpose, Hermitian transpose, determinant, rank, and column space, whereas $\|\mathbf{a}\|$ and $\|\mathbf{A}\|_F$ denote the Euclidean norm of a vector $\mathbf{a}$ and the Frobenius norm of a matrix $\mathbf{A}$, respectively. The notation $\mathbf{I}_n$ refers to the $n \times n$ identity matrix, and $\mathbf{0}$ denotes the all-zero vector or matrix, where the subscript is omitted when the dimension is evident from the context. The operators $\operatorname{vec}(\cdot)$, $\otimes$, $\langle\cdot,\cdot\rangle$, and $\Re\{\cdot\}$ correspond to the vectorization, Kronecker product, inner product, and the real part of a complex argument, respectively. In terms of matrix inequalities, $\mathbf{A}\succeq\mathbf{B}$ (or $\mathbf{A} \succ \mathbf{B}$) implies that $\mathbf{A}-\mathbf{B}$ is PSD (or PD). Finally, $\mathbb{E}_\mathbf{X}[\cdot]$ denotes the expectation with respect to the random variables $\mathbf{X}$, and $\mathcal{CN}(\mathbf{0},\mathbf{R})$ denotes the zero-mean circularly symmetric complex Gaussian distribution with covariance matrix $\mathbf{R}$, whose probability density function is denoted by $p_{\mathcal{CN}}(\mathbf{x};\mathbf{R})$.

\section{System Model and Problem Formulation}

We consider a MIMO system in which sensing signals are transmitted from $N_t$ transmit antennas and the echo signals are received by $N_r$ receive antennas. We use $T$ snapshots for sensing, each representing an independent observation. The target response matrix is denoted by ${\bf{H}} \in \mathbb{C}^{N_t \times N_r}$.

To detect the presence or absence of a target, we formulate a binary hypothesis testing problem based on the observed sensing signal returns. The hypotheses are given by:
\begin{align}
& \CMcal{H}_0: \mathbf{Y} = \mathbf{X}\mathbf{C}_0 + \mathbf{N}, \label{h0}\\
& \CMcal{H}_1: \mathbf{Y} = \mathbf{X}\mathbf{H} + \mathbf{X}\mathbf{C}_1 +  \mathbf{N}, \label{h1}
\end{align}
where $\mathbf{Y} \in \mathbb{C}^{T \times N_r}$ is the observation matrix collected over $T$ snapshots across $N_r$ receive antennas, $\mathbf{X} \in \mathbb{C}^{T \times N_t}$ represents the transmitted signal matrix and $\mathbf{N}\in \mathbb{C}^{T \times N_r}$  is the additive noise matrix. The matrices $\mathbf{C}_0$ and $ \mathbf{C}_1$ denote the clutter responses under $\CMcal{H}_0$ and $\CMcal{H}_1$, respectively. This formulation accommodates general scenarios by allowing for distinct clutter responses ($\mathbf{C}_0\neq\mathbf{C}_1$). Nevertheless, the proposed method remains applicable without modification for the special case of identical clutter responses, where $\mathbf{C}_0=\mathbf{C}_1$.\\

Assuming that the receive antennas are sufficiently separated, the columns of the random matrices $\mathbf{H}$, $\mathbf{C}_0$, $\mathbf{C}_1$, and $\mathbf{N}$ are modeled as mutually independent and identically distributed (i.i.d.) zero-mean complex Gaussian vectors. 
The covariance matrices $\mathbf{R}_H$, $\mathbf{R}_0$, $\mathbf{R}_1$, and $\mathbf{R}_N$ represent the spatial statistics of the target response, clutter under $\CMcal{H}_0$, clutter under $\CMcal{H}_1$, and noise, respectively. We note that this assumption follows the standard modeling practice in MIMO radar literature \cite{haimovich2007mimo, de2007design}.

The primary objective in the hypothesis testing \eqref{h0}, \eqref{h1} is to reliably distinguish between the absence ($\CMcal{H}_0$) and presence ($\CMcal{H}_1$) of the target. According to the Neyman–Pearson criterion \cite{van2004detection}, the optimal detector is the likelihood-ratio test (LRT), given by:
\begin{align}
\Lambda(\mathbf{Y}) = \frac{p(\mathbf{Y}|\CMcal{H}_1)}{p(\mathbf{Y}|\CMcal{H}_0)} \underset{\CMcal{H}_0}{\overset{\CMcal{H}_1}{\gtrless}} \eta,
\label{eq:lrt}
\end{align}
where $\eta$ is a threshold chosen based on the desired probability of false alarm or detection, and $p(\mathbf{Y}|\CMcal{H}_1)$ and $p(\mathbf{Y}|\CMcal{H}_0)$ denote the likelihood functions under the respective hypotheses. These likelihood functions are given by:
\begin{align}
    p(\mathbf{Y}|\CMcal{H}_i) &= \frac{1}{(\pi^T|\mathbf{K}_i|)^{N_r}} \exp\left(-\bar{\mathbf{y}}^{\sf H}(\mathbf{I}_{N_r}\otimes\mathbf{K}_i)^{-1}\bar{\mathbf{y}}\right),\quad i=0,1,
    \label{lrt}
\end{align}
where 
\begin{align}
    & \bar{\mathbf{y}} = \operatorname{vec}(\mathbf{Y})\in\mathbb{C}^{N_rT}, \\
    & \mathbf{K}_0=\mathbf{X}\mathbf{R}_0\mathbf{X}^{\sf H}+\mathbf{R}_N, \label{eq:K0} \\   
    & \mathbf{K}_1=\mathbf{X}\mathbf{R}_{H1}\mathbf{X}^{\sf H}+\mathbf{R}_N, \label{eq:K1}\\
    & \mathbf{R}_{H1}=\mathbf{R}_H+\mathbf{R}_1. \label{RH1}
\end{align}
%for $i=0,1$.
% $\mathbf{K}_0=\mathbf{X}\mathbf{R}_0\mathbf{X}^{\sf H}+\mathbf{R}_N$,  $\mathbf{K}_1=\mathbf{X}\mathbf{R}_{H1}\mathbf{X}^{\sf H}+\mathbf{R}_N$, 
    % $\mathbf{R}_{H1}=\mathbf{R}_H+\mathbf{R}_1$, and  
% In \eqref{lrt}, we have $\bar{\mathbf{y}}\in\mathbb{C}^{N_rT}$. 
    % with $\bar{\mathbf{y}}\in\mathbb{C}^{N_rT}$ 

The asymptotic performance of the LRT is characterized by the Chernoff–Stein lemma, which relates the exponential decay rate of the type-II error (miss-detection) probability to the KLD divergence \cite{coverEIT}. Specifically, letting $\alpha$ denote the fixed type-I error (false-alarm) probability, the minimum achievable type-II error probability, denoted by $\beta_T(\alpha)$, decays as characterized by
% \begin{align}
%     \beta_T(\alpha) = \mathop{\inf}_{\alpha(\CMcal{D}) \le \alpha} \beta(\CMcal{D}),
% \end{align}
% then the type-II error (miss-detection) probability decays as characterized by
\begin{align}
% \lim_{T \rightarrow \infty} \frac{1}{T} \log \beta_T = -D_{\text{KL}}(p(\mathbf{Y}|\CMcal{H}_0) \,\|\, p(\mathbf{Y}|\CMcal{H}_1)), \label{stein}
\lim_{T \rightarrow \infty} \frac{1}{T} \log \beta_T(\alpha) = -D_{\text{KL}}({\bf{Y}}_0 \Vert {\bf{Y}}_1), \label{stein}
\end{align}
where $D_{\text{KL}}(\cdot \Vert \cdot)$ denotes the KLD.
To be specific, the KLD $D_{\text{KL}}({\bf{Y}}_0 \Vert {\bf{Y}}_1)$ is defined as
\begin{align}
    D_{\text{KL}}(\mathbf{Y}_0 \Vert \mathbf{Y}_1)
    &= -\mathbb{E}_{\mathbf{Y}|\CMcal{H}_0} \left[\log \frac{p(\mathbf{Y}|\CMcal{H}_1)}{p(\mathbf{Y}|\CMcal{H}_0)} \right]  \nonumber \\
    &= N_r \left( \log \left|\mathbf{K}_0^{-1} \mathbf{K}_1 \right| + \operatorname{Tr} (\mathbf{K}_1^{-1} \mathbf{K}_0 ) \right) - N_r T, \label{eq:kld}
\end{align}
where $\mathbf{Y}_0$ and $\mathbf{Y}_1$ denote the signals received under $\CMcal{H}_0$ and $\CMcal{H}_1$, respectively. 
As such, when the number of snapshots is sufficiently large (i.e., $T > \max\{N_r, N_t\}$), 
in order to reduce the type-II error, it is desirable to design the transmit waveform $\mathbf{X}$ such that
% this result motivates the design of the transmit waveform $\mathbf{X}$ to maximize the KL divergence.
\begin{align}
    \underset{\mathbf{X}}{\text{maximize}}\quad &D_{\text{KL}}(\mathbf{Y}_0\Vert \mathbf{Y}_1), \label{opt_main}\\
    \text{subject to}\quad & \operatorname{Tr}(\mathbf{X}\mathbf{X}^{\sf H}) \leq P_t. \nonumber
\end{align}
% where $\mathbf{Y}_0$ and $\mathbf{Y}_1$ denote the signals received under $\CMcal{H}_0$ and $\CMcal{H}_1$, respectively, and 
Here, $P_t$ is the transmit power constraint. 
This paper primarily focuses on addressing \eqref{opt_main}. 
% Addressing \eqref{opt_main} is our main target 

Ignoring a constant term, the objective of the considered problem \eqref{opt_main} can be written as
\begin{align}
% D_{\text{KL}}(\mathbf{Y}_0 \,\|\, \mathbf{Y}_1)
f({\bf{X}})
= \log|\mathbf{K}_0^{-1}\mathbf{K}_1|
+\operatorname{Tr}(\mathbf{K}_1^{-1}\mathbf{K}_0). \label{eq:obj_raw}
\end{align}

% response 2.2
\begin{remark}[Scope of the Gaussian KLD formulation]\label{rem:gaussian_kld}
\normalfont
    The closed-form KLD objective in \eqref{eq:kld} is specific to the zero-mean complex Gaussian observation model, in which the waveform affects the objective only through the covariance matrices $\mathbf{K}_0(\mathbf{X})$ and $\mathbf{K}_1(\mathbf{X})$; the proposed FP/MM reformulations exploit precisely this structure, consistent with standard detection- and information-theoretic radar/sensing waveform design~\cite{kay2007optimal,chen2009mimo,zhu2017information,tang:tsp:10,tang2015relative,tang:tsp:18}.
    For general non-Gaussian observations, the KLD is given by
    \begin{align}
        D_{\mathrm{KL}}(p_0\|p_1)
        =
        \int p_0(\mathbf{Y};\mathbf{X})
        \log
        \frac{p_0(\mathbf{Y};\mathbf{X})}{p_1(\mathbf{Y};\mathbf{X})}
        \, d\mathbf{Y},
    \end{align}
    which depends on the full likelihoods $p_0$ and $p_1$ rather than on second-order statistics
    alone, so a universal closed form with the same log-det/trace structure is unavailable. This case requires a model-specific treatment and is beyond the scope of this paper. It is left for future work.
    % Consequently, a universal waveform-dependent closed form with the same log-determinant/trace structure as in the Gaussian case is not generally available. Extending the framework to non-Gaussian settings would thus require a model-specific divergence, error exponent, or robust surrogate together with a separate optimization analysis, which is beyond the scope of this paper and is left for future work.
\end{remark}

This Gaussian KLD objective depends on the design variable $\mathbf{X}$ through
$\mathbf{K}_0=\mathbf{X}\mathbf{R}_0\mathbf{X}^{\sf H}+\mathbf{R}_N$ and
$\mathbf{K}_1=\mathbf{X}\mathbf{R}_{H1}\mathbf{X}^{\sf H}+\mathbf{R}_N$,
which are quadratic in $\mathbf{X}$.  
% $\log|\cdot|$ and $\operatorname{Tr}(\cdot^{-1}\cdot)$ is 
Thus, concavity/convexity of the log-det and trace functions is lost when they are composed with $\mathbf{X}$.  
In particular, the coupling term 
$\operatorname{Tr}(\mathbf{K}_1^{-1}\mathbf{K}_0)$ introduces a 
quadratic dependence on $\mathbf{X}$, making the Hessian indefinite.  
For instance, even considering the scalar case, the objective function has the form of 
\begin{align}
f(x)=\log\frac{b x^2+\sigma^2}{a x^2+\sigma^2}
+\frac{a x^2+\sigma^2}{b x^2+\sigma^2},    
\end{align}
which is neither convex nor concave in $x$. Together with practical power or structure constraints, the problem becomes a highly nonconvex program, for which typical optimization methods are not directly applicable. 
In the next section, we address this by proposing a FP-based optimization technique. 

\section{Proposed FP-based KLD Optimization}\label{sec:FP-KLD}

In this section, we present our main method to solve \eqref{opt_main}. 
We begin by rewriting the objective function \eqref{eq:obj_raw} as 
\begin{align}
f(\mathbf{X})
&= \log\bigl|\mathbf{K}_0^{-1}\mathbf{K}_1\bigr| \;+\;\operatorname{Tr}( \mathbf{K}_1^{-1}\mathbf{K}_0 ) \nonumber\\
&= \log\bigl|\mathbf{I}_{T} + \mathbf{K}_0^{-1}\mathbf{X}(\mathbf{R}_{H1}-\mathbf{R}_0)\mathbf{X}^{\sf H}\bigr|\nonumber\\
   &\quad\quad\quad- \operatorname{Tr}(\mathbf{K}_1^{-1}\mathbf{X}(\mathbf{R}_{H1}-\mathbf{R}_0)\mathbf{X}^{\sf H}).\label{eq:objective}
\end{align}

% Without loss of generality, we assume $\mathbf{R}_{H1}-\mathbf{R}_0 \succ 0$ \cite{kay1993fundamentals}. 
% This is a natural condition in binary detection problems: under $\CMcal{H}_1$, 
% an additional signal component is present compared to $\CMcal{H}_0$, which 
% necessarily enlarges the covariance matrix. 
% If this condition did not hold, the two hypotheses would not be distinguishable, and the detection 
% problem would be ill-posed. In particular, if the clutter response is identical under both hypotheses (${\bf{C}}_0 = {\bf{C}}_1$), we have 
% \begin{align}
% \mathbf{R}_{H1}-\mathbf{R}_0 = \mathbf{R}_H \succ 0,    
% \end{align}
% thereby the condition holds naturally. 
Without loss of generality, we assume $\bfR_{H1}-\bfR_0\succeq \mathbf{0}$. This condition naturally arises in binary detection problems: under $\CMcal{H}_1$, an additional signal component is present compared to $\CMcal{H}_0$, which enlarges the covariance matrix. In particular, in the typical case where the clutter responses are identical under both hypotheses ($\bfC_0=\bfC_1$), we have $\bfR_{H1}-\bfR_0=\bfR_H\succ \mathbf{0}$, and the condition holds naturally when $\bfR_H$ is full-rank in practical sensing scenarios. When $\bfR_{H1}-\bfR_0$ is strictly PD, it admits the standard Cholesky factorization $\bfR_{H1}-\bfR_0=\bfL\bfL^\herm$ with $\bfL\in\bbC^{N_t\times N_t}$. In the more general PSD case where $\operatorname{rank}(\bfR_{H1}-\bfR_0)=r\le N_t$, we instead employ the thin factorization $\bfR_{H1}-\bfR_0=\bfL\bfL^\herm$ with $\bfL\in\bbC^{N_t\times r}$ obtained via the truncated eigen-decomposition. All subsequent derivations in this section remain valid with the dimensions of the auxiliary variables adjusted accordingly. The only case beyond the scope of this work is when $\bfR_{H1}-\bfR_0$ is indefinite, which corresponds to a non-physical situation where the detection problem itself is ill-posed. Such a case is not covered in this work and would require a separate derivation.

Under this assumption, we write $\mathbf{R}_{H1}-\mathbf{R}_0 = \mathbf{L}\mathbf{L}^{\sf H}$, using the Cholesky factorization in the PD case and the thin factorization in the PSD case.
The objective becomes
\begin{align}
f(\mathbf{X})
&= \log\bigl|\mathbf{I}_{N_t} + (\mathbf{X}\mathbf{L})^{\sf H}\mathbf{K}_0^{-1}(\mathbf{X}\mathbf{L})\bigr|
   - \operatorname{Tr}((\mathbf{X}\mathbf{L})^{\sf H}\mathbf{K}_1^{-1}(\mathbf{X}\mathbf{L})). \label{eq:obj_re}
\end{align}
Nevertheless, \eqref{eq:obj_re} contains matrix-ratio terms inside both the log-det and the trace. To address this challenge, we introduce auxiliary variables and apply a sequence of matrix transformations, each leading to a tractable surrogate. 
We note that our approach is rooted in the principles of matrix FP 
\cite{shen2019optimization, shen2024accelerating, chen2025fast}.
%, which can be naturally interpreted within the MM framework. 
We provide further clarification regarding the connection between the applied FP and the MM framework in Remark \ref{rem:mm}.
% We discuss this in more details in Remark \ref{rem_mm}. 
% In each iteration, the transformation yields a surrogate objective that is tractable to optimize, thereby reducing the original problem to a sequence of simpler subproblems. 

% has the matrix-ratio forms inside both the log-det and the trace. We therefore introduce auxiliary variables and apply a sequence of matrix transforms, each yielding a tractable surrogate. This approach is rooted in the principles of matrix FP \cite{shen2019optimization, chen2025fast} and can be interpreted within the minorization-maximization (MM) framework. Each transform constructs a surrogate function that is more amenable to optimization, allowing us to iteratively solve a sequence of simpler subproblems.

% {\textbf{Matrix Lagrangian dual transform}}: 
First, to handle the challenging log-det term in \eqref{eq:obj_re}, we apply the matrix Lagrangian dual transform \cite{shen2019optimization}. This transform provides a tight and global lower bound on the log-det function. 
\begin{lemma}[Matrix Lagrangian dual transform {\cite{shen2019optimization}}] \label{lem:trans}
Let $\mathbf{Q},\mathbf{\Gamma}\succeq \mathbf{0}$ and define
$\mathbf{Z}=\mathbf{I}+\mathbf{Q}\succ \mathbf{0}$, 
$\mathbf{U}=\mathbf{I}+\mathbf{\Gamma}\succ \mathbf{0}$. Then
\begin{align}
\log|\mathbf{I}+\mathbf{Q}|
\;\ge\;
\log|\mathbf{I}+\mathbf{\Gamma}| 
\;+\; \operatorname{Tr}(\mathbf{I})
\;-\; \operatorname{Tr}((\mathbf{I}+\mathbf{\Gamma})(\mathbf{I}+\mathbf{Q})^{-1}),\label{eq:lemma1}
\end{align}
with equality if and only if $\mathbf{Q}=\mathbf{\Gamma}$.
\end{lemma}
\begin{proof}[Proof sketch]
The inequality follows from the basic bound
\begin{align}
\log|\mathbf{M}| \le \operatorname{Tr}(\mathbf{M}) - \operatorname{Tr}(\mathbf{I}) \quad (\mathbf{M}\succ \mathbf{0}),   \end{align}
with equality if and only if $\mathbf{M}=\mathbf{I}$. 
When $\mathbf{M}=\mathbf{U}\mathbf{Z}^{-1}$, we have 
\begin{align}
\log|\mathbf{U}\mathbf{Z}^{-1}| \le \operatorname{Tr}(\mathbf{U}\mathbf{Z}^{-1}) - \operatorname{Tr}(\mathbf{I}).    
\end{align}
Rearranging gives
\begin{align}
\log|\mathbf{Z}| \ge \log|\mathbf{U}| + \operatorname{Tr}(\mathbf{I}) - \operatorname{Tr}(\mathbf{U}\mathbf{Z}^{-1}).  \end{align}
Substituting $\mathbf{Z}=\mathbf{I}+\mathbf{Q}$, $\mathbf{U}=\mathbf{I}+\mathbf{\Gamma}$ completes the proof.
% yields the claimed result (with $n=T=\dim(\mathbf{Z})$). 
% Equality holds iff $\mathbf{U}\mathbf{Z}^{-1}=\mathbf{I}$, i.e., $\mathbf{U}=\mathbf{Z}$, 
% which is equivalent to $\mathbf{\Gamma}=\mathbf{Q}$.
\end{proof}
% \begin{align}
%     \log \left|\mathbf{I}+\mathbf{Q}\right| \geq \log\left|\mathbf{I}+\mathbf{\Gamma}\right|+\operatorname{Tr}(\mathbf{I})-\operatorname{Tr}((\mathbf{I}+\mathbf{\Gamma})(\mathbf{I}+\mathbf{Q})^{-1}) \label{dual_trans}
% \end{align}
% which holds for any positive semi-definite matrices $\mathbf{Q}$ and $\mathbf{\Gamma}$. 
% We note that the equality holds if and only if $\mathbf{Q}=\mathbf{\Gamma}$. 

% We leverage this principle to construct a surrogate for the first term of our objective funtion (15), $\log\left| \mathbf{I}_{N_t} + (\mathbf{X}\mathbf{L})^{\sf H}\mathbf{K}_0^{-1}(\mathbf{X}\mathbf{L})\right|$. To do this, we identify the matrix $\mathbf{Q}$ in the inequality (16) with our term $(\mathbf{X}\mathbf{L})^{\sf H}\mathbf{K}_0^{-1}(\mathbf{X}\mathbf{L})$ which yields a new lower bound
Applying Lemma \ref{lem:trans} to the first term in \eqref{eq:obj_re}, we get 
\begin{align}
    &\log\bigl|\mathbf{I}_{N_t} + (\mathbf{X}\mathbf{L})^{\sf H}\mathbf{K}_0^{-1}(\mathbf{X}\mathbf{L})\bigr| \nonumber\\
    &\geq \log \left| \mathbf{I}_{N_t}+\mathbf{\Gamma} \right| +\operatorname{Tr}(\mathbf{I}_{N_t})\nonumber\\
    &\quad - \operatorname{Tr}( (\mathbf{I}_{N_t}+\mathbf{\Gamma})(\mathbf{I}_{N_t}+ (\mathbf{X}\mathbf{L})^{\sf H}\mathbf{K}_0^{-1}(\mathbf{X}\mathbf{L}))^{-1}).\label{eq:dual_trans_ineq}
\end{align}
By modifying the final trace term of inequality \eqref{eq:dual_trans_ineq} using the Woodbury matrix identity \cite{horn2012matrix} and combining the second term from \eqref{eq:obj_re}, we obtain an equivalent reformulation of the problem. 
% the problem can be reformulated.
This introduces an auxiliary variable $\mathbf{\Gamma}\in \mathbb{H}_{+}^{N_t}$ and yields a new surrogate for the original objective function, resulting in the formulation in \eqref{eq:reform_dual}.
\begin{align}
\underset{\mathbf{X},\mathbf{\Gamma}}{\text{maximize}} \quad & f_\ell(\mathbf{X},\mathbf{\Gamma})\label{eq:reform_dual} \\
\text{subject to} \quad & \operatorname{Tr}(\mathbf{X}\mathbf{X}^{\sf H}) \le P_t, \nonumber
\end{align}
where
\begin{align}
f_\ell(\mathbf{X},\mathbf{\Gamma})
= \log\bigl|\mathbf{I}_{N_t} + \mathbf{\Gamma}\bigr| - \operatorname{Tr}(\mathbf{\Gamma})
  + \operatorname{Tr} (\mathbf{\Gamma}\,(\mathbf{X}\mathbf{L})^{\sf H}\mathbf{K}_1^{-1}(\mathbf{X}\mathbf{L}) ).\label{eq:surr_1}
\end{align}
This transformation effectively decouples the matrix inverse from the log-det operator, resulting in a more tractable structure. 
% The key to the MM algorithm is to choose the 
% Especially, by finding a proper auxiliary variable $\mathbf{\Gamma}$ at each iteration, we reach the surrogate function $f_\ell$ corresponding to a tight lower bound of the original objective at the current point as presented in Lemma \ref{lem:trans}. 
In particular, by selecting an appropriate auxiliary variable $\mathbf{\Gamma}$ at each iteration, we obtain a suitable surrogate function $f_\ell$, which achieves a tight lower bound of the original objective at the current point as stated in Lemma~\ref{lem:trans}. 
% Specifically, this tight bound is achieved when:
% \begin{align}
%     \mathbf{\Gamma}^{\star} = (\mathbf{X}\mathbf{L})^{\sf H}\mathbf{K}_0^{-1}(\mathbf{X}\mathbf{L}).\label{eq:opt_Gamma}
% \end{align}
% % $$. 
% Substituting $\mathbf{\Gamma}^\star$ back into $f_\ell(\mathbf{X},\mathbf{\Gamma})$ recovers a function equivalent to the original objective with respect to $\mathbf{X}$, thus ensuring the validity of this transformation.
Specifically, by its equality condition, this tight bound is achieved when:
\begin{align}
    \mathbf{\Gamma}^{\star} = (\mathbf{X}\mathbf{L})^{\sf H}\mathbf{K}_0^{-1}(\mathbf{X}\mathbf{L}).\label{eq:opt_Gamma}
\end{align}
% Now we focus on the transformed objective function $f_{\ell}$ in \eqref{eq:surr_1}. 
% Unfortunately, the objective function $f_\ell$ still contains the term $\operatorname{Tr}(\mathbf{\Gamma}\left(\mathbf{X}\mathbf{L}\right)^{\sf H}\mathbf{K}_1^{-1}\left(\mathbf{X}\mathbf{L}\right))$, 
% which is challenging to handle due to the matrix inverse. 
% To deal with this, we further apply the \emph{matrix quadratic transform}. This transform is specified in the following lemma. 
However, $f_\ell$ in \eqref{eq:surr_1} remains intractable due to the matrix inverse in the trace term. To deal with this, we further apply the following lemma.

\begin{lemma}[Matrix quadratic transform \cite{shen2019optimization}]\label{lem:quad_trans}
    Let $\mathbf{S} \in \mathbb{H}_+$ and $\mathbf{D}\in \mathbb{H}_{++}$.
    Then for any $\mathbf{\Psi}$,
    \begin{align}
        \operatorname{Tr}(\sqrt{\mathbf{S}}^{\sf H}\mathbf{D}^{-1}\sqrt{\mathbf{S}})\geq 2\Re\{\operatorname{Tr}(\mathbf{\Psi}^{\sf H}\sqrt{\mathbf{S}})\} - \operatorname{Tr}(\mathbf{\Psi}^{\sf H}\mathbf{D}\mathbf{\Psi}),\label{eq:qt_ineq}
    \end{align}
    where $\mathbf{S} = \sqrt{\mathbf{S}}\sqrt{\mathbf{S}}^{\sf H}$, with equality if and only if $\mathbf{\Psi}=\mathbf{D}^{-1}\sqrt{\mathbf{S}}$. 
\end{lemma}
\begin{proof}[Proof sketch]
    Consider the matrix $\mathbf{Z}$ defined as
    \begin{align}
        \mathbf{Z} = (\mathbf{\Psi}-\mathbf{D}^{-1}\sqrt{\mathbf{S}})^{\sf H}\mathbf{D}(\mathbf{\Psi}-\mathbf{D}^{-1}\sqrt{\mathbf{S}})
    \end{align}
    Since $\mathbf{D}$ is PD, $\mathbf{Z}$ is a PSD matrix.
    \begin{align}
        \mathbf{Z} &= (\mathbf{\Psi}-\mathbf{D}^{-1}\sqrt{\mathbf{S}})^{\sf H}\mathbf{D}(\mathbf{\Psi}-\mathbf{D}^{-1}\sqrt{\mathbf{S}}) \nonumber \\
        &=\mathbf{\Psi}^{\sf H}\mathbf{D}\mathbf{\Psi}-2\Re\{\mathbf{\Psi}^{\sf H}\sqrt{\mathbf{S}}\}+\sqrt{\mathbf{S}}^{\sf H}\mathbf{D}^{-1}\sqrt{\mathbf{S}} \succeq \mathbf{0}
    \end{align}
    with equality if and only if $\mathbf{\Psi}-\mathbf{D}^{-1}\sqrt{\mathbf{S}}=\mathbf{0}$. 
    % As $\mathbf{D}$ is PSD, this is equivalent to $\mathbf{\Psi}=\mathbf{D}^{-1}\sqrt{\mathbf{S}}$. Thus, inequality \eqref{eq:qt_ineq} holds for any $\mathbf{\Psi}$.
\end{proof}

 % Lemma \ref{lem:quad_trans} allows us to replace the matrix inverse term with an equivalent optimization problem that is quadratic in a new auxiliary variable $\mathbf{\Psi}\in\mathbb{C}^{T\times N_t}$. Applying this principle yields our final surrogate objective:

Lemma~\ref{lem:quad_trans} enables us to reformulate the matrix inverse term 
as an equivalent quadratic optimization with a new auxiliary variable 
$\mathbf{\Psi}\in\mathbb{C}^{T\times N_t}$. 
By applying this transformation, we arrive at the final surrogate objective:
\begin{align}
\underset{\mathbf{X},\mathbf{\Gamma},\mathbf{\Psi}}{\text{maximize}} \quad & f_q(\mathbf{X},\mathbf{\Gamma},\mathbf{\Psi}) \\
\text{subject to} \quad & \operatorname{Tr}(\mathbf{X}\mathbf{X}^{\sf H}) \le P_t, \nonumber
\end{align}
where
\begin{align}
f_q(\mathbf{X},\mathbf{\Gamma},\mathbf{\Psi})
&= \log\bigl|\mathbf{I}_{N_t} + \mathbf{\Gamma}\bigr| - \operatorname{Tr}(\mathbf{\Gamma}) \nonumber\\
&\quad + \operatorname{Tr}(2\,\Re\{(\mathbf{X}\mathbf{L})^{\sf H}\mathbf{\Psi}\,\mathbf{\Gamma}\}-\mathbf{\Psi}^{\sf H}  \mathbf{K}_1\mathbf{\Psi}\,\mathbf{\Gamma})\nonumber\\
&= \underbrace{\log|\mathbf{I}_{N_t} + \mathbf{\Gamma}| - \operatorname{Tr}(\mathbf{\Gamma})
 - \operatorname{Tr}(\mathbf{\Psi}^{\sf H}\mathbf{R}_N\mathbf{\Psi}\mathbf{\Gamma})}_{\text{constant in }\mathbf{X}} \nonumber\\
&\quad + 2\,\Re\big\{\operatorname{Tr}(\mathbf{\Gamma}\mathbf{\Psi}^{\sf H}\mathbf{X}\mathbf{L})\big\}
- \operatorname{Tr}(\mathbf{X}\mathbf{R}_{H1}\mathbf{X}^{\sf H}\mathbf{\Psi}\mathbf{\Gamma}\mathbf{\Psi}^{\sf H}).\label{eq:obj_qt}
\end{align}
We note that the optimal choice for the auxiliary variable $\mathbf{\Psi}$ that satisfies the equality in Lemma \ref{lem:quad_trans} is:
\begin{align}
    \mathbf{\Psi}^\star = \mathbf{K}_1^{-1}\mathbf{X}\,\mathbf{L}. \label{eq:opt_Y}
\end{align}
With $\mathbf{\Gamma}$ and $\mathbf{\Psi}$ fixed at the corresponding iteration, the dependence on $\mathbf{X}$ arises through the last two terms. 
% The linear term $2\,\Re\{\operatorname{Tr}(({\bf{XL}})^{\sf H} {\bf{\Psi}} {\bf{\Gamma}})\}$ is affine in $\mathbf{X}$. Second, by substituting  $\mathbf{K}_1=\mathbf{X}\mathbf{R}_{H1}\mathbf{X}^{\sf H}+\mathbf{R}_N$,
% we derive 
% % Expanding $\mathbf{K}_1=\mathbf{X}\mathbf{R}_{H1}\mathbf{X}^{\sf H}+\mathbf{R}_N$ in \eqref{eq:obj_qt} with fixed $(\mathbf{\Gamma},\mathbf{\Psi})$, we obtain
% which leads to 
Notably, since $\mathbf{R}_{H1}$ and $ \mathbf{\Psi}\mathbf{\Gamma}\mathbf{\Psi}^{\sf H}$ are both PSD, 
$f_q$ is a concave quadratic form with respect to $\mathbf{X}$; 
thus the global optimal solution under the convex constraint can be obtained by solving the Karush-Kuhn-Tucker (KKT) conditions of the associated Lagrangian \cite{boyd2004convex}. 
Specifically, we introduce a Lagrange multiplier $\mu\geq0$ for the power constraint and define the Lagrangian function as:
\begin{align}
    \mathcal{L}(\mathbf{X},\mu)=f_q(\mathbf{X},\mathbf{\Gamma},\mathbf{\Psi})-\mu(\operatorname{Tr}(\mathbf{X}\mathbf{X}^{\sf H})-P_t).
\end{align}
By deriving the stationary condition, i.e., setting the gradient of $\mathcal{L}$ to zero, we obtain a Sylvester-type equation \cite{sylvester1, sylvester2}, given by: 
\begin{align}
    \mathbf{A}\,\mathbf{X}\,\mathbf{R}_{H1} + \mu\,\mathbf{X} = \mathbf{B},\label{eq:sylvester}
\end{align}
where $\mathbf{A} = \mathbf{\Psi}\,\mathbf{\Gamma}\,\mathbf{\Psi}^{\sf H}$, $\mathbf{B} = \mathbf{\Psi}\,\mathbf{\Gamma}\,\mathbf{L}^{\sf H}$.
% Analyzing the KKT conditions for the $\mathbf{X}$-subproblem provides a direct analytical solution in the form of a Sylvester-type equation \cite{sylvester1, sylvester2}:
% \begin{align}
%     \mathbf{A}\,\mathbf{X}\,\mathbf{R}_{H1} + \mu\,\mathbf{X} = \mathbf{B},\label{eq:sylvester}
% \end{align}
% where $\mathbf{A} = \mathbf{\Psi}\,\mathbf{\Gamma}\,\mathbf{\Psi}^{\sf H}$, $\mathbf{B} = \mathbf{\Psi}\,\mathbf{\Gamma}\,\mathbf{L}^{\sf H}$, and $\mu$ is the Lagrange multiplier. 
\eqref{eq:sylvester} can be solved by vectorizing it into a standard linear system: $( \bar{\mathbf{A}} + \mu\,\mathbf{I}_{N_tT})\,\bar{\mathbf{x}} = \bar{\mathbf{b}}$,
with $\bar{\mathbf{x}} = \operatorname{vec}(\mathbf{X})$, $\bar{\mathbf{b}} = \operatorname{vec}(\mathbf{B})$, and $\bar{\mathbf{A}} = \mathbf{R}_{H1}^{\sf T} \otimes \mathbf{A}$, which typically requires operations with a complexity of $\mathcal{O}((N_tT)^3)$. 
Since $\bar{\mathbf{A}}$ is the Kronecker product of two PSD matrices, it is inherently PSD. Consequently, for any $\mu>0$, the matrix $\bar{\mathbf{A}}+\mu\mathbf{I}_{N_tT}$ becomes PD, guaranteeing the existence of a unique solution $\bar{\mathbf{x}}=(\bar{\mathbf{A}}+\mu\mathbf{I}_{N_tT})^{-1}\bar{\mathbf{b}}$. 
We refer to this method as FP-KLD and summarize the overall procedure in Algorithm~\ref{alg:FP-KLD}.

Beyond the algorithmic description, we present several discussions regarding the proposed FP-KLD to elucidate the theoretical foundations of the algorithm. 
Specifically, we rigorously analyze its convergence properties by establishing a connection to the MM framework and provide an intuitive interpretation through block coordinate ascent.  

\begin{algorithm}[t]
\caption{{FP-KLD} (Matrix FP + Quadratic Transform)}\label{alg:FP-KLD}
\textbf{Input}:$P_t$,$\tau>0$, $t_{\max}$, 
$\mathbf{R}_{H1},\mathbf{R}_0,\mathbf{R}_N$, and $\mathbf{X}_{\text{init}}$ \\
\textbf{Precompute}: $\mathbf{R}_{H1}-\mathbf{R}_0=\mathbf{L}\mathbf{L}^{\sf H}$ \\
\textbf{Initialize}: $\mathbf{X}^{(0)}\gets \mathbf{X}_{\text{init}}$; set $t\gets 0$
\\
\While{$t<t_{\max}$}{
   Compute $\mathbf{K}_0$ by \eqref{eq:K0} and $\mathbf{K}_1$ by \eqref{eq:K1}.\\  
  Update $\mathbf{\Gamma}^{(t+1)}$ by \eqref{eq:opt_Gamma}.\\
  Update $\mathbf{\Psi}^{(t+1)}$ by \eqref{eq:opt_Y}.\\
  Update $\mathbf{X}^{(t+1)}$ by solving \eqref{eq:sylvester}\\
  \If{$\displaystyle \frac{f(\mathbf{X}^{(t+1)})-f(\mathbf{X}^{(t)})}{f(\mathbf{X}^{(t)})}<\tau$}{\textbf{break}}
  $t \leftarrow t+1$
}
\textbf{Output}: $\mathbf{X}^{\star}\leftarrow \mathbf{X}^{(t+1)}$.
\end{algorithm}

\begin{remark}[MM interpretation] \normalfont \label{rem:mm}

We note that the proposed FP-based optimization mechanism can also be understood within the MM framework. 
Specifically, as shown in Lemma \ref{lem:trans}, for any given $\mathbf X$, introducing the auxiliary variable 
$\mathbf\Gamma\succeq 0$ produces the surrogate function \eqref{eq:surr_1}, which serves as a global lower bound of the original objective $f(\mathbf X)$. 
Crucially, the bound becomes tight at the point $\mathbf\Gamma^\star(\mathbf X)=(\mathbf X\mathbf L)^{\sf H}\mathbf K_0^{-1}(\mathbf X\mathbf L)$, which indicates $\max_{\mathbf\Gamma} f_\ell(\mathbf X,\mathbf\Gamma)=f(\mathbf X)$. 
Similar to this, in Lemma \ref{lem:quad_trans}, we introduce an additional auxiliary variable $\mathbf{\Psi}$, the quadratic surrogate \eqref{eq:obj_qt} is obtained, which is tight at $\mathbf{\Psi}^\star(\mathbf X)=\mathbf K_1^{-1}\mathbf X\mathbf L$. This implies $\max_{\mathbf{\Psi}} f_q(\mathbf X,\mathbf\Gamma,\mathbf{\Psi})=f_\ell(\mathbf X,\mathbf\Gamma)$.

For this reason, the iteration that sets $\mathbf\Gamma^{(t)}=\mathbf\Gamma^\star(\mathbf X^{(t)})$ and 
$\mathbf{\Psi}^{(t)}=\mathbf{\Psi}^\star(\mathbf X^{(t)})$ defines the surrogate
\begin{align}
g(\mathbf X\,|\,\mathbf X^{(t)}) \triangleq
f_q\bigl(\mathbf X,\mathbf\Gamma^{(t)},\mathbf{\Psi}^{(t)}\bigr),
\end{align}
which satisfies two key MM properties: (a) it is a global lower bound of the original objective, and 
(b) it is tight at the current iterate $\mathbf X^{(t)}$. 
Consequently, the MM principle guarantees the monotonic improvement 
\begin{align}
f(\mathbf X^{(t+1)}) \ge
g(\mathbf X^{(t+1)}|\mathbf X^{(t)}) \ge
g(\mathbf X^{(t)}|\mathbf X^{(t)}) = f(\mathbf X^{(t)}).\label{eq:ineq_tightness}
\end{align}
Since the feasible set $\mathcal X=\{\mathbf{X}:\operatorname{Tr}(\mathbf{X}\mathbf{X}^{\sf H})\leq P_t\}$ is compact and $f$ is continuous, the sequence $\{f(\mathbf{X}^{(t)}) \}$ converges to a finite limit, and any accumulation point of $\mathbf{X}^{(t)}$ is a stationary point of the original problem.
We clarify that the connection between matrix FP and MM was well explained in \cite{shen2019optimization}. 
\end{remark}

\begin{remark}[BCA interpretation] \normalfont
The FP iteration can alternatively be interpreted from the perspective of block coordinate ascent (BCA). 
Define the lifted objective
\begin{align}
F(\mathbf X,\mathbf\Gamma,\mathbf{\Psi})\triangleq f_q(\mathbf X,\mathbf\Gamma,\mathbf{\Psi}),
\end{align}
which by construction satisfies 
\begin{align}
f(\mathbf X)=\max_{\mathbf\Gamma\succeq 0,\;\mathbf{\Psi}} F(\mathbf X,\mathbf\Gamma,\mathbf{\Psi}).
\end{align}
In this view, the variables are partitioned into three blocks: $(\mathbf{\Psi})$, $(\mathbf\Gamma)$, and $(\mathbf X)$. 
Maximizing $F$ with respect to one block while holding the others fixed yields the following alternating updates:
\begin{align}
\mathbf{\Gamma}^{(t+1)}&=\arg\max_{\mathbf{\Gamma}}F(\mathbf{X}^{(t)},\mathbf{\Gamma}, \mathbf{\Psi}^{(t)})=(\mathbf X^{(t)}\mathbf L)^{\sf H}\mathbf K_0^{-1}(\mathbf X^{(t)}\mathbf L)\\
\mathbf{\Psi}^{(t+1)} &= \arg\max_{\mathbf{\Psi}}\;F(\mathbf X^{(t)},\mathbf\Gamma^{(t+1)},\mathbf{\Psi})
= \mathbf K_1^{-1}\mathbf X^{(t)}\mathbf L,\\[2pt]
\mathbf X^{(t+1)} &= \arg\max_{\mathbf X\in\mathcal X}\;F(\mathbf X,\mathbf\Gamma^{(t+1)},\mathbf{\Psi}^{(t+1)}),
\end{align}
%where $\mathcal X$ denotes the feasible set imposed by system constraints (e.g., power budgets or shaping). 
From the above update processes, two observations follow. 
First, the updates for $\mathbf{\Gamma}$ and $\mathbf{\Psi}$ admit closed-form solutions, 
which can be computed efficiently at each iteration. 
Second, the $\mathbf X$-update reduces to a structured convex or quadratic optimization problem, 
for which efficient numerical solvers or even analytic updates may be available. 
Because each block is optimized exactly, the lifted objective $F$ is monotonically non-decreasing across iterations. 
Furthermore, since $F$ is a tight reformulation of $f$, the monotonicity of $F$ directly implies monotonicity of 
the original KLD objective $f$. 

For this reason, the FP can be equivalently viewed as a BCA on the augmented problem, guaranteeing convergence under mild regularity conditions.
\end{remark}

\section{Accelerated MM-based KLD Optimization}\label{sec:A-MM-KLD}

In the previous section, we develop a tractable iterative algorithm to solve a generic KLD optimization problem. 
Nonetheless, its final subproblem, as shown in \eqref{eq:sylvester}, requires solving a large-scale linear system at each iteration. This step, originating from the matrix inversion embedded in the quadratic transform, incurs a high computational complexity of order $\mathcal{O}((N_tT)^3)$, which can be prohibitive in practice. 
To resolve this issue, we present reduced-complexity FP-KLD variants in this section, referred to as MM-KLD and A-MM-KLD. The two key ideas underlying these methods are as follows.
First, to eliminate the costly subproblem, we apply a \emph{nonhomogeneous relaxation} technique to the concave quadratic surrogate $f_q$. This relaxation replaces the complex, anisotropic curvature of the subproblem, i.e., the term $\bar{\mathbf{x}}^{\sf H}\bar{\mathbf{A}}\bar{\mathbf{x}}$, with a simple isotropic spectral bound of the form $\lambda||\bar{\mathbf{x}}||^2$, yielding an update for $\mathbf{X}$ in a simple, closed form.

However, this computational gain comes at the cost of an increased number of iterations required for convergence. 
This is because the nonhomogeneous bound is inherently looser than the original quadratic surrogate, causing the algorithm to take more conservative steps and thus increasing the total number of iterations. 
To counteract this, our second key idea is to employ acceleration techniques. By interpreting the iterative algorithm as a fixed-point mapping, we apply acceleration techniques such as the Steffensen-type method (STEM) method to substantially reduce the number of iterations, resulting in a computationally light and fast-converging algorithm \cite{SQUAREM}.

\subsection{Nonhomogeneous Relaxation}\label{subsec:relaxation}
Algorithm \ref{alg:FP-KLD} provides a direct method for solving the $\mathbf{X}$-subproblem by analyzing the KKT conditions. In this section, we present an approach that further simplifies the subproblem to yield an update for $\mathbf{X}$ in a closed-form expression, avoiding the need to solve a complicated linear system and to compute a matrix inversion.

To achieve this, we focus on the vectorized quadratic term within the objective function. As previously shown, the parts of $f_q$ involving $\mathbf{X}$ can be written as:
\begin{align}
    &2\,\Re\big\{\operatorname{Tr}(\mathbf{\Gamma}\mathbf{\Psi}^{\sf H}\mathbf{X}\mathbf{L})\big\}
    - \operatorname{Tr}(\mathbf{X}\mathbf{R}_{H1}\mathbf{X}^{\sf H}\mathbf{\Psi}\mathbf{\Gamma}\mathbf{\Psi}^{\sf H})\nonumber\\
    &\ = \operatorname{Tr}(2\Re\{(\mathbf{X}\mathbf{L})^{\sf H}\mathbf{\Psi}\mathbf{\Gamma}\}) - \operatorname{Tr}(\mathbf{X}\mathbf{R}_{H1}\mathbf{X}^{\sf H}\mathbf{A}) \nonumber\\
    &\ =2\Re\{\bar{\mathbf{x}}^{\sf H}\bar{\mathbf{b}}\}-\bar{\mathbf{x}}^{\sf H}\bar{\mathbf{A}}\bar{\mathbf{x}},
\end{align}
where $\bar{\mathbf{x}}=\operatorname{vec}(\mathbf{X})$. The core of this alternative approach lies in constructing a simpler surrogate for the quadratic term $\bar{\mathbf{x}}^{\sf H}\bar{\mathbf{A}}\bar{\mathbf{x}}$ by applying a nonhomogeneous relaxation. The key to this relaxation is the following lemma.
\begin{lemma}[Nonhomogeneous bound \cite{sun2016majorization}]\label{lem:nonhomogeneous}
    Consider two Hermitian matrices $\mathbf{P}$ and $\mathbf{G}$ such that $\mathbf{P} \prec \mathbf{G}$. 
    Then, for any $\mathbf{x}$ and $\mathbf{z}$,
    \begin{align}
        \mathbf{x}^{\sf H}\mathbf{P}\mathbf{x}
        \le \mathbf{x}^{\sf H}\mathbf{G}\mathbf{x}
          + 2\,\Re\{\mathbf{x}^{\sf H}(\mathbf{P}-\mathbf{G})\mathbf{z}\}
          + \mathbf{z}^{\sf H}(\mathbf{G}-\mathbf{P})\mathbf{z},\label{eq:nonhomogeneous_bound}
    \end{align}
    with equality if and only if $\mathbf{z}=\mathbf{x}$.
\end{lemma}
\begin{proof}[Proof sketch]
    Since $\mathbf{G}-\mathbf{P}\succ\mathbf{0}$, $\mathbf{v}^H(\mathbf{G}-\mathbf{P})\mathbf{v}\geq 0$ for any $\mathbf{v}$. Let $\mathbf{v}=\mathbf{z}-\mathbf{x}$, then
    \begin{align}
        \mathbf{v}^{\sf H}(\mathbf{G}-\mathbf{P})\mathbf{v}&=\left(\mathbf{z}-\mathbf{x}\right)^{\sf H}(\mathbf{G}-\mathbf{P})\left(\mathbf{z}-\mathbf{x}\right)\nonumber\\
        &=\mathbf{z}^{\sf H}(\mathbf{G}-\mathbf{P})\mathbf{z}-2\Re\{\mathbf{x}^{\sf H}(\mathbf{G}-\mathbf{P})\mathbf{z}\}\nonumber\\
        &\qquad+\mathbf{x}^{\sf H}\mathbf{G}\mathbf{x}-\mathbf{x}^{\sf H}\mathbf{P}\mathbf{x}\geq 0
    \end{align}
    Thus, \eqref{eq:nonhomogeneous_bound} holds for any $\mathbf{x}$ and $\mathbf{z}$. The equality holds \textit{if and only if} $\mathbf{z}=\mathbf{x}$.
\end{proof}

Applying Lemma~\ref{lem:nonhomogeneous} to the quadratic term 
$\bar{\mathbf{x}}^{\sf H}\bar{\mathbf{A}}\bar{\mathbf{x}}$ with 
$\mathbf{P}=\bar{\mathbf{A}}$ and $\mathbf{G}=\bar\lambda\,\mathbf{I}_{N_t T}$, where 
$\lambda_p$ is the largest eigenvalue of $\bar{\mathbf{A}}$ and 
$\bar\lambda\triangleq\lambda_p+\delta$ for a small $\delta>0$ (so that 
$\bar{\mathbf{A}}\prec \bar\lambda\,\mathbf{I}_{N_t T}$), we introduce an auxiliary 
variable $\bar{\mathbf{z}}\in\mathbb{C}^{N_tT}$ and derive the surrogate
\begin{align}
\underset{\mathbf{X},\mathbf{\Gamma},\mathbf{\Psi},\bar{\mathbf{z}}}{\text{maximize}}\quad 
& f_h(\bar{\mathbf{x}},\mathbf{\Gamma},\mathbf{\Psi},\bar{\mathbf{z}}) \\
\text{subject to}\quad & \|\bar{\mathbf{x}}\|^2 \le P_t,\quad 
\bar{\mathbf{x}}=\operatorname{vec}(\mathbf{X}) \nonumber
\end{align}
with
\begin{align}
f_h(\bar{\mathbf{x}},\mathbf{\Gamma},\mathbf{\Psi},\bar{\mathbf{z}}) \label{eq:f_h}
&= \log\bigl|\mathbf{I}_{N_t} + \mathbf{\Gamma}\bigr| - \operatorname{Tr}(\mathbf{\Gamma})
  - \operatorname{Tr}(\mathbf{R}_N\mathbf{A}) 
  + 2\,\Re\bigl\{\bar{\mathbf{x}}^{\sf H}\bar{\mathbf{b}}\bigr\} \nonumber\\
&\quad + 2\,\Re\bigl\{\bar{\mathbf{x}}^{\sf H}(\bar\lambda\,\mathbf{I}_{N_t T}-\bar{\mathbf{A}})\,\bar{\mathbf{z}}\bigr\}
  + \bar{\mathbf{z}}^{\sf H}(\bar{\mathbf{A}}-\bar\lambda\,\mathbf{I}_{N_t T})\,\bar{\mathbf{z}}\nonumber\\
  &\qquad\qquad\qquad\qquad\qquad\qquad- \bar\lambda\,\bar{\mathbf{x}}^{\sf H}\bar{\mathbf{x}}\,.
\end{align}
Given $\mathbf{X}$ (or equivalently $\bar{\mathbf{x}}$), by the equality condition of Lemma \ref{lem:nonhomogeneous}, the optimal value of the new auxiliary variable simply follows as
\begin{align}
    \bar{\mathbf{z}}^\star= \bar{\mathbf{x}}\label{eq:opt_z}
\end{align}
Subsequently, for fixed $\{\mathbf{\Gamma},\mathbf{\Psi},\bar{\mathbf{z}}\}$ at current iteration, we update $\bar{\mathbf{x}}$ by maximizing the objective $f_h$ of \eqref{eq:f_h}. Since $f_h$ is a concave quadratic function with respect to $\bar{\mathbf{x}}$, its unconstrained maximizer is found by setting the gradient with respect to $\bar{\mathbf{x}}$ to zero:
\begin{align}
    2\left(\bar{\mathbf{b}} + (\bar{\lambda}\mathbf{I}_{N_t T}-\bar{\mathbf{A}})\bar{\mathbf{z}}-\bar{\lambda}\bar{\mathbf{x}}\right)=\mathbf{0}
\end{align}

The final update for $\bar{\mathbf{x}}^\star$ is obtained by projecting this unconstrained solution onto the feasible set defined by the power constraint, which results in the closed-form expression:
\begin{align}
    \bar{\mathbf{x}}^\star
    = \sqrt{P_t}\,\frac{\bar{\mathbf{b}} + (\bar{\lambda}\mathbf{I}_{N_t T}-\bar{\mathbf{A}})\bar{\mathbf{z}}}
                  {\|\bar{\mathbf{b}} + (\bar{\lambda}\mathbf{I}_{N_t T}-\bar{\mathbf{A}})\bar{\mathbf{z}}\|}.\label{eq:update_x_nonhomo}
\end{align}

Based on these derived update rules, the overall optimization strategy is established. This three-stage framework--dual, quadratic, and nonhomogeneous transforms--yields a sequence of convex subproblems in $\mathbf{X}$ that can be solved efficiently until convergence. Specifically, the inclusion of the final nonhomogeneous transform replaces the direct solution of a linear system, which has a complexity of $\mathcal{O}((N_tT)^3)$, with a simple closed-form update. The dominant cost is computing the spectral radius $\bar{\lambda}$ via $k$ power iterations, requiring only $\mathcal{O}(k(N_tT)^2)$, where $k$ is typically a small number \cite{golub2013matrix, press2007numerical}. The complete iterative procedure incorporating the nonhomogeneous relaxation is formally summarized in Algorithm~\ref{alg:MM-KLD}, which we refer to as MM-KLD.

We now justify the convergence of the proposed MM-KLD by showing it adheres to the MM principle. The nonhomogeneous relaxation introduces a second level of minorization. We define a new surrogate function for the relaxed algorithm as 
\begin{align}
    h(\mathbf{X}|\mathbf{X}^{(t)})\triangleq f_h(\bar{\mathbf{x}},\mathbf{\Gamma}^{(t+1)}, \mathbf{\Psi}^{(t+1)},\bar{\mathbf{z}}^{(t+1)}), 
\end{align}
% \begin{align}
% \end{align}
where all the auxiliary variables $\mathbf{\Gamma}^{(t+1)}, \mathbf{\Psi}^{(t+1)},\bar{\mathbf{z}}^{(t+1)}$ are updated based on $\mathbf{X}^{(t)}$.

This new surrogate $h(\mathbf{X}|\mathbf{X}^{(t)})$ satisfies the two key properties of an MM algorithm. First, by construction, $f_h$ is a global lower bound of $f_q$ (Lemma \ref{lem:nonhomogeneous}), and $f_q$ is a global lower bound of $f$, which by transitivity makes $h(\mathbf{X}|\mathbf{X}^{(t)})$ a global lower bound of $f(\mathbf{X})$. Second, at the current iterate $\mathbf{X}^{(t)}$, all auxiliary variables are chosen precisely to meet the equality conditions of their respective transformations. These choices for $\mathbf{\Gamma}$ and $\mathbf{\Psi}$ ensure the tightness of the first-level surrogate $g$ as shown in Remark \ref{rem:mm}, while the choice $\bar{\mathbf{z}}^{(t+1)}=\bar{\mathbf{x}}^{(t)}$ ensures the tightness of the second-level surrogate $h$. 

Maximizing this tight surrogate yields the update $\mathbf{X}^{(t+1)}$, leading to the standard ascent inequality, analogous to \eqref{eq:ineq_tightness}:
\begin{align}
    f(\mathbf X^{(t+1)}) \ge
h(\mathbf X^{(t+1)}|\mathbf X^{(t)}) \ge
h(\mathbf X^{(t)}|\mathbf X^{(t)}) = f(\mathbf X^{(t)}).\label{eq:ineq_tightness_2}
\end{align}
Thus, the inequality chain \eqref{eq:ineq_tightness_2} guarantees a monotonic ascent for the original objective $f(\mathbf{X})$. This confirms that Algorithm \ref{alg:MM-KLD} is also a valid instantiation of the MM algorithm.

Although the nonhomogeneous relaxation is tight at the current iterate $\mathbf{X}^{(t)}$, \eqref{eq:f_h} replaces the anisotropic curvature $\bar{\mathbf{x}}^{\sf H}\bar{\mathbf{A}}\bar{\mathbf{x}}$ with an isotropic spectral bound $\bar{\lambda}\bar{\mathbf{x}}^{\sf H}\bar{\mathbf{x}}$ along with corresponding linear terms, leading to a more conservative step.
Therefore, FP-KLD produces a steeper ascent per iteration than the nonhomogeneous variant, whereas in terms of runtime, the nonhomogeneous method is significantly faster due to its much lower per-iteration cost.

\begin{algorithm}[t]
\caption{MM-KLD}\label{alg:MM-KLD}
\textbf{Input}:$P_t$, $\tau>0$, $t_{\max}$, $\delta>0$, 
$\mathbf{R}_{H1},\mathbf{R}_0,\mathbf{R}_N$, and $\mathbf{X}_{\text{init}}$ \\
\textbf{Precompute}: $\mathbf{R}_{H1}-\mathbf{R}_0=\mathbf{L}\mathbf{L}^{\sf H}$ \\
\textbf{Initialize}: $\mathbf{X}^{(0)}\gets \mathbf{X}_{\text{init}}$; set $t\gets 0$
\\
\While{$t<t_{\max}$}{
   Compute $\mathbf{K}_0$ by \eqref{eq:K0} and $\mathbf{K}_1$ by \eqref{eq:K1}\\
  
  Update $\mathbf{\Gamma}^{(t+1)}$ by \eqref{eq:opt_Gamma}.\\
  Update $\mathbf{\Psi}^{(t+1)}$ by \eqref{eq:opt_Y}.\\
  Update $\bar{\mathbf{z}}^{(t+1)}$ by \eqref{eq:opt_z}\\
  Update $\mathbf{X}^{(t+1)}$ by solving \eqref{eq:update_x_nonhomo} and reshaping $\bar{\mathbf{x}}$.\\
  \If{$\displaystyle \frac{f(\mathbf{X}^{(t+1)})-f(\mathbf{X}^{(t)})}{f(\mathbf{X}^{(t)})}<\tau$}{\textbf{break}}
  $t \leftarrow t+1$
}
\textbf{Output}: $\mathbf{X}^{\star}\leftarrow \mathbf{X}^{(t+1)}$.
\end{algorithm}

\begin{remark}[Comparison with AWD-MM]\normalfont
    It is instructive to compare our proposed framework with AWD-MM, i.e., the 
    MM-based adaptive waveform-design baseline of Tang \textit{et al.}~\cite{tang:tsp:18}. 
    AWD-MM adopts a piecewise decomposition approach: it splits the intricate KLD 
    objective into several log-det and trace terms and then employs first-order 
    Taylor expansions, or supporting hyperplanes, to bound the convex parts based 
    on local gradient information. The final quadratic surrogate is then formed by 
    summing these local approximations.
    
    In contrast, our FP-based approach reformulates the objective globally using 
    auxiliary variables. Instead of relying on gradient-based linearizations, we 
    apply a sequence of matrix FP transformations to alter the algebraic structure 
    of the problem. This procedure decouples the matrix inverses from the design 
    variables within the log-det and trace-ratio terms, naturally yielding a 
    quadratic surrogate.
    
    This difference leads to a significant computational advantage. In AWD-MM, 
    updating the waveform requires solving a large-scale linear system at every 
    iteration. Our A-MM-KLD avoids this bottleneck by incorporating the 
    nonhomogeneous relaxation into the FP-derived quadratic surrogate. By replacing 
    the anisotropic curvature with an isotropic spectral bound, we obtain a 
    closed-form update and avoid solving the large-scale linear system.
\end{remark}

\subsection{Acceleration}

In the previous subsection, we introduced the nonhomogeneous relaxation (Algorithm \ref{alg:MM-KLD}) to reduce the high per-iteration cost of the original FP-KLD algorithm. This computational gain, however, stems from using the isotropic surrogate $f_h$, which is inherently looser than the anisotropic quadratic surrogate $f_q$. Consequently, while each step is much faster, the algorithm typically requires a larger total number of iterations to reach the same solution. To compensate for this trade-off, we use an acceleration technique.

To motivate the use of acceleration schemes, it is essential to interpret our optimization framework as a fixed-point mapping. 
At iteration $t$, the auxiliary variables $\mathbf{\Gamma}^{(t)}$, $\mathbf{\Psi}^{(t)}$, and the vectorized auxiliary $\bar{\mathbf{z}}^{(t)}$ are updated in closed form based on the current $\mathbf{X}^{(t)}$. Subsequently, the next iterate, $\mathbf{X}^{(t+1)}$, is obtained by maximizing the resulting surrogate objective, which is now parameterized by these fixed auxiliary variables. This leads to the following remark. 

\begin{remark}[Fixed-point iteration interpretation]\normalfont
The two-stage procedure, where the auxiliary variables are first determined by $\mathbf{X}^{(t)}$ and then in turn determine $\mathbf{X}^{(t+1)}$, can be encapsulated by a single nonlinear mapping operator, $\CMcal{M}(\cdot)$. This allows the entire update rule to be expressed concisely as a fixed-point iteration:
\begin{align}
    \mathbf{X}^{(t+1)} = \CMcal{M}(\mathbf{X}^{(t)})
\end{align}
The algorithm's goal is thus to find the fixed point $\mathbf{X}^\star$ that satisfies $\mathbf{X}^\star=\CMcal{M}(\mathbf{X}^\star)$. This interpretation is particularly valuable because algorithms derived from the MM principle, while guaranteeing monotonic convergence, often exhibit a slow, linear convergence rate if the problem is ill-conditioned \cite{hunter2004tutorial, lange1999numerical}. This slow convergence can make achieving a high-precision solution prohibitively time-consuming.
By framing our method as a fixed-point iteration, we can directly leverage a suite of well-established acceleration schemes designed for general fixed-point iterations to enhance practical performance.
\end{remark}

Applying acceleration modifies the update step of MM-KLD. Instead of simply setting $\mathbf{X}^{(t+1)}=\CMcal{M}(\mathbf{X}^{(t)})$, we use the unaccelerated output $\CMcal{M}(\mathbf{X}^{(t)})$ as an input to an acceleration formula, which then computes the final $\mathbf{X}^{(t+1)}$ for the next iterate.

While various well-established acceleration schemes exist, such as Polyak's heavy-ball \cite{polyak1964some} and Nesterov acceleration \cite{nesterov1983method}, we focus on STEM \cite{SQUAREM} for our framework. 
STEM is a modern vector generalization of the classical scalar Steffensen's method \cite{steffensen1933remarks, nievergelt1991aitken}, designed to accelerate general fixed-point iterations.
This method is an effective derivative-free technique that requires only the fixed-point mapping $\CMcal{M}(\cdot)$ itself, yet it constructs a secant approximation that achieves a substantially faster practical convergence near the solution, offering a notable speed-up over the linear convergence of the baseline MM algorithm \cite{hunter2004tutorial}.

The core of STEM is to define a residual function:
\begin{align}
    \mathbf{R}(\mathbf{X})= \CMcal{M}(\mathbf{X})-\mathbf{X},
\end{align}
and, at iteration $t$, compute the two successive images
\begin{align}
\mathbf{\Theta}_1 &= \CMcal{M}(\mathbf{X}^{(t)}),\label{eq:Steffensen_1}\\
\mathbf{\Theta}_2 &= \CMcal{M}(\mathbf{\Theta}_1).\label{eq:Steffensen_2}
\end{align}
Let the first residual and the second one be
\begin{align}
    \mathbf{\Delta}_t &= \mathbf{\Theta}_1 - \mathbf{X}^{(t)} = \mathbf{R}(\mathbf{X}^{(t)})\label{eq:residual_1}\\
    \mathbf{W}_t &= \mathbf{\Theta}_2 - 2\mathbf{\Theta}_1 +\mathbf{X}^{(t)}=\mathbf{R}(\mathbf{\Theta}_1)-\mathbf{R}(\mathbf{X}^{(t)})\label{eq:residual_2}
\end{align}
The accelerated candidate is taken along the residual direction with a step size $\gamma_t$ as
\begin{align}
    \gamma_t &= \frac{\langle \mathbf{\Delta}_t, \mathbf{\Delta}_t \rangle}{\langle \mathbf{\Delta}_t, \mathbf{W}_t \rangle}\label{eq:step_size}\\
    \mathbf{X}^{(t+1)}_{\text{cand}}&=\mathbf{X}^{(t)}-\gamma_t\mathbf{\Delta}_t\label{eq:X_cand}\\
    \mathbf{X}^{(t+1)} &= \CMcal{P}(\mathbf{X}^{(t+1)}_{\text{cand}})
\end{align}
Here $\CMcal{P}(\cdot)$ is the projection operator that maps a candidate solution back onto the constraint feasible set. Intuitively, $\gamma_t$ is chosen so that the next residual is orthogonal to the current residual under a local secant approximation, thereby removing the leading error component.

Because STEM steps do not inherently guarantee the non-decreasing property, we employ a simple monotonicity check with backtracking: if the objective fails to increase, we shrink the step length via $\gamma_t\leftarrow(\gamma_t-1)/2$, recompute the candidate, and repeat until acceptance. As this rule is iterated, $\gamma_t\to -1$, in which case the update falls back to $\mathbf{X}^{(t)}-\gamma_t\mathbf{\Delta}_t=\mathbf{\Theta}_1=\CMcal{M}(\mathbf{X}^{(t)})$, which is exactly the baseline MM step. This mechanism ensures that the algorithm preserves the monotonicity guaranteed by the original MM framework.

\begin{remark}[Why Steffensen-type acceleration?]\normalfont
    Momentum-based methods such as Polyak's and Nesterov's improve the constant of linear convergence but remain first-order and do not elevate the order of convergence. STEM is a fundamentally different approach: it is a derivative-free, secant-based scheme inspired by Newton-type iterations. Instead of relying on past directions, it probes the local geometry of the map $\mathcal{M}(\cdot)$ by computing two successive images. As shown in \eqref{eq:residual_2}, the difference $\mathbf{W}_{t}$ implicitly captures information about the local behavior of the map. STEM uses this information to build a secant model and take an approximate Newton-type step, which in our experiments yields markedly faster empirical convergence near the solution than the baseline MM iteration.

    This makes it a synergistic choice for our framework. The primary cost of STEM is the second map evaluation. However, we have just demonstrated the MM-KLD, whose entire purpose was to make the single iterate computationally cheap by avoiding the linear system solve of size $N_tT$.
\end{remark}

Algorithm \ref{alg:A-MM-KLD} formalizes this procedure, integrating the computationally efficient map from Algorithm \ref{alg:MM-KLD} with the STEM's acceleration steps. We name this complete framework A-MM-KLD. This approach combines the low per-iteration cost of the nonhomogeneous relaxation with the fast practical convergence of STEM, resulting in a highly efficient and robust algorithm for KLD maximization.

\begin{algorithm}[t]
\caption{A-MM-KLD}\label{alg:A-MM-KLD}
\textbf{Input}:$P_t$, $\tau>0$, $t_{\max}$, $\delta>0$, 
$\mathbf{R}_{H1},\mathbf{R}_0,\mathbf{R}_N$, and $\mathbf{X}_{\text{init}}$ \\
\textbf{Precompute}: $\mathbf{R}_{H1}-\mathbf{R}_0=\mathbf{L}\mathbf{L}^{\sf H}$ \\
\textbf{Initialize}: $\mathbf{X}^{(0)}\gets \mathbf{X}_{\text{init}}$; set $t\gets 0$
\\
\While{$t<t_{\max}$}{
  Compute $\mathbf{K}_0$ by \eqref{eq:K0} and $\mathbf{K}_1$ by \eqref{eq:K1}\\
  Compute $\CMcal{M}(\mathbf{X}^{(t)}), \CMcal{M}(\mathbf{\Theta}_1)$ using \eqref{eq:opt_Gamma}, \eqref{eq:opt_Y}, \eqref{eq:opt_z} and solving \eqref{eq:update_x_nonhomo} followed by reshaping.\\
  Compute the residuals $\mathbf{\Delta}_t, \mathbf{W}_t$ by \eqref{eq:residual_1}, \eqref{eq:residual_2}. \\
  Compute the step size $\gamma_t$ by \eqref{eq:step_size}. \\
  \Repeat{$f(\mathbf{X}^{(t+1)}) \ge f(\mathbf{X}^{(t)})$}{
   $\mathbf{X}^{(t+1)}_{\text{cand}}\leftarrow \mathbf{X}^{(t)}-\gamma_t\mathbf{\Delta}_t$\\
   $\mathbf{X}^{(t+1)}\leftarrow \sqrt{P_t}\frac{\mathbf{X}^{(t+1)}_{\text{cand}}}{||\mathbf{X}^{(t+1)}_{\text{cand}}||_F}$\\
   \If{$f(\mathbf{X}^{(t+1)}) < f(\mathbf{X}^{(t)})$}{
     $\gamma_t\leftarrow(\gamma_t-1)/2$
   }
 }  
  \If{$\displaystyle \frac{f(\mathbf{X}^{(t+1)})-f(\mathbf{X}^{(t)})}{f(\mathbf{X}^{(t)})}<\tau$}{\textbf{break}}
  $t \leftarrow t+1$
}
\textbf{Output}: $\mathbf{X}^{\star}\leftarrow \mathbf{X}^{(t+1)}$.
\end{algorithm}

\section{Extensions and Application Scenarios}\label{sec:applications}

Having established a general KLD optimization framework, we now demonstrate its
versatility through two representative applications. These applications illustrate that once the Gaussian KLD is rendered tractable, the same FP/MM machinery applies without modification to problems in which KLD appears as part of the objective. We first show that the proposed FP-KLD can be seamlessly extended to the joint waveform design in ISAC systems. Subsequently, we apply the framework to a multiple random access scenario to maximize activity detection reliability.

\subsection{Joint Waveform Design for ISAC}\label{subsec:isac}
The proposed FP-KLD, originally developed for KLD maximization in sensing, can be directly extended to joint waveform design in ISAC systems. This subsection explains its applicability to ISAC waveform design.

A primary objective in ISAC is to optimize a joint utility function that jointly accounts for sensing and communication performance metrics. One possible formulation is a weighted sum of KLD (as a sensing performance metric) and the MI (as a communication performance metric), given by:
\begin{align}
    \underset{\mathbf{X}}{\text{maximize}}\quad & (1-\rho)D_{\text{KL}}(\mathbf{Y}_0\Vert \mathbf{Y}_1)+\rho I(\mathbf{X};\mathbf{Y}_{\text{c}}),  \label{eq:isac_objective}\\
\text{subject to} \quad & \operatorname{Tr}(\mathbf{X}\mathbf{X}^{\sf H}) \le P_t, 
   \nonumber
\end{align}
where $\mathbf{Y}_c=\mathbf{H}_c\mathbf{X}^{\sf H}+\mathbf{N}_c$. Here, $\mathbf{Y}_c$, $\mathbf{H}_c$, and $\mathbf{N}_c$ denote the received signal, channel matrix, and zero-mean Gaussian noise with covariance $\mathbf{R}_{n,c}$, respectively.
This formulation corresponds to a standard approach for characterizing the fundamental trade-off between sensing and communication, consistent with recent ISAC frameworks \cite{chiriyath2017radar, al2023unified, fei2024revealing}. 
As such, it provides a unified performance metric for joint waveform design. 
As discussed in the earlier section, a significant challenge in solving \eqref{eq:isac_objective} lies in the disparity among the available optimization tools. 
Our proposed FP-KLD directly addresses this bottleneck by providing a computationally tractable surrogate for the KLD. 
Specifically, as FP is a well-established tool for MI maximization in communication systems, 
our approach renders the ISAC optimization problem amenable to FP; resulting in that the composite ISAC objective function is efficiently handled by applying the same FP principles \cite{shenFPtsp1, shen2019optimization, chen2025fast}.

Now we explain the detailed mechanism for joint ISAC waveform design using FP. 
By applying the determinant lemma \cite{horn2012matrix}, the MI is given by 
\begin{align}
I(\mathbf{X};\mathbf{Y}_c)&=\log|\mathbf{I}+\mathbf{R}_{n,c}^{-1}\mathbf{H}_c\mathbf{X}^{\sf H}\mathbf{X}\mathbf{H}_c^{\sf H}|\nonumber\\
    &=\log|\mathbf{I}+(\mathbf{H}_c\mathbf{X}^{\sf H})^{\sf H}\mathbf{R}_{n,c}^{-1}\mathbf{H}_c\mathbf{X}^{\sf H}|
    \label{eq:comm_mi}
\end{align}
To  unify this with the KLD optimization, we apply the proposed sequence of transforms introducing two auxiliary variables, as in FP-KLD. First, Lemma~\ref{lem:trans} with an auxiliary variable $\mathbf{\Gamma}_c\in\mathbb{H}_+^{T}$ yields a lower bound
\begin{align}
    f_{c,l}(\mathbf{X}, \mathbf{\Gamma}_c)&=\log|\mathbf{I}+\mathbf{\Gamma}_c|-\operatorname{Tr}(\mathbf{\Gamma}_c)+\operatorname{Tr}((\mathbf{I}+\mathbf{\Gamma}_c)\cdot\nonumber\\&
    (\mathbf{H}_c\mathbf{X}^{\sf H})^{\sf H}(\mathbf{H}_c\mathbf{X}^{\sf H}\mathbf{X}\mathbf{H}_c^{\sf H}+\mathbf{R}_{n,c})^{-1}(\mathbf{H}_c\mathbf{X}^{\sf H})).
\end{align}
Subsequently, applying Lemma~\ref{lem:quad_trans} with an auxiliary variable $\mathbf{\Psi}_c\in\mathbb{C}^{N_c\times T}$ results in the final surrogate as:
\begin{align}
    f_{c,q}(\mathbf{X}, \mathbf{\Gamma}_c, &\mathbf{\Psi}_c) =\log|\mathbf{I}+\mathbf{\Gamma}_c|-\operatorname{Tr}(\mathbf{\Gamma}_c)+\operatorname{Tr}((\mathbf{I}+\mathbf{\Gamma}_c)\cdot\nonumber\\
    &(2\Re\{(\mathbf{H}_c\mathbf{X}^{\sf H})^{\sf H}\mathbf{\Psi}_c\}-\mathbf{\Psi}_c^{\sf H}(\mathbf{H}_c\mathbf{X}^{\sf H}\mathbf{X}\mathbf{H}_c^{\sf H}+\mathbf{R}_{n,c})\mathbf{\Psi}_c))\nonumber\\
    &\qquad=\mathrm{const.} + 2\Re\{\operatorname{Tr}((\mathbf{I}+\mathbf{\Gamma}_c)\mathbf{\Psi}_c^{\sf H}\mathbf{H}_c\mathbf{X}^{\sf H})\}\nonumber\\
    &\qquad\quad-\operatorname{Tr}(\mathbf{X}\mathbf{H}_c^{\sf H}\mathbf{\Psi}_c(\mathbf{I}+\mathbf{\Gamma}_c)\mathbf{\Psi}_c^{\sf H}\mathbf{H}_c\mathbf{X}^{\sf H} ),
\end{align}
where $\mathrm{const.}$ denotes the terms independent of $\mathbf{X}$. 
Since $f_{c,q}$ is a concave quadratic function of $\mathbf{X}$, sharing the exact algebraic structure as in \eqref{eq:obj_qt}, the joint ISAC problem \eqref{eq:isac_objective} is efficiently solved by iteratively maximizing the global surrogate $(1-\rho)f_q+\rho f_{c,q}$. This involves updating the auxiliary variables $\mathbf{\Gamma}$, $\mathbf{\Psi}$, $\mathbf{\Gamma}_c$, $\mathbf{\Psi}_c$, followed by a unified quadratic update for $\mathbf{X}$. 
This highlights a seamless extension of the proposed FP-KLD method to joint waveform optimization in ISAC systems.

\subsection{Multiple Random Access}\label{subsec:random_access}
In this subsection, we present another application scenario for the proposed optimization technique. To be specific, we consider a multiple random access problem, which aims to design a random access waveform that maximizes the reliability of activity detection.

In typical cellular networks, random access is handled by assigning quasi-orthogonal sequences (e.g., Zadoff-Chu \cite{chu1972polyphase, frank1962phase}) to devices, which simplifies detection. However, the performance of such fixed sequences degrades severely in the presence of strong multi-user interference (MUI) or in non-orthogonal settings that may arise from an increasing number of users. 
To resolve this, we instead formulate the problem as maximizing the sum of KLD between the active and inactive hypotheses for all users, averaged over the activity patterns of interfering devices. This approach allows us to jointly optimize the waveforms to be robust to MUI, offering a significant advantage over fixed-sequence designs.

\subsubsection{Model and Problem Formulation}
We consider a random access setting with $K$ devices. Over $T$ snapshots, the received signal $\mathbf{Y}\in\mathbb{C}^{T\times N_r}$ is
\begin{align}
    \mathbf{Y} \;=\; \sum_{i=1}^K \mathbf{X}_i \mathbf{H}_i \;+\; \mathbf{N},
\end{align}
where $\mathbf{X}_i\in \mathbb{C}^{T\times N_t}$ is the waveform of device $i$. The columns of $\mathbf{H}_i\in \mathbb{C}^{N_t\times N_r}$ are i.i.d., each distributed as $\mathcal{CN}(\mathbf{0}, \mathbf{R}_i)$. Thus, the covariance matrix of $\operatorname{vec}(\mathbf{H}_i)$ is $\mathbf{I}_{N_r}\otimes\mathbf{R}_i$. The noise $\mathbf{N}$ is zero-mean complex Gaussian with independent columns, each with covariance $\mathbf{R}_N$; hence, the covariance matrix of $\operatorname{vec}(\mathbf{N})$ is equal to $\mathbf{I}_{N_r}\otimes \mathbf{R}_N$.

For each device $i$, we test in parallel
\begin{align}
    \CMcal{H}_0^{(i)}&:\ \text{device $i$ is inactive}\nonumber\\
    \CMcal{H}_1^{(i)}&:\ \text{device $i$ is active}.\nonumber
\end{align}
Let $\mathbf{s}_{-i} \in \{0,1\}^{K-1}$ denote the on/off activity pattern of the interfering devices (i.e., all devices except device $i$), and let $\CMcal{S}_{-i}$ be the set of all such patterns. We assume the per-user activity priors $\{p_j\}_{j=1}^K$ are known and independent across devices. Then the probability mass function of $\mathbf{s}_{-i}$ is
\begin{align}
w(\mathbf{s}_{-i}) \;=\; \prod_{j\neq i} p_j^{\,s_j}\,(1-p_j)^{\,1-s_j},\qquad \mathbf{s}_{-i}\in\CMcal{S}_{-i},
\end{align}
where $s_j$ is the element of $\mathbf{s}_{-i}$ corresponding to device $j$. Crucially, the distribution of the interference pattern $\mathbf{s}_{-i}$ is independent of the hypothesis tested for device $i$. Conditioned on $\mathbf{s}_{-i}$, the observation is complex Gaussian with zero-mean and covariance matrices
\begin{align}
    \mathbf{K}_{i,0}(\mathbf{s}_{-i}) &= \mathbf{R}_N + \sum_{j\neq i} s_j\mathbf{X}_j \mathbf{R}_j \mathbf{X}_j^{\sf H},\\
    \mathbf{K}_{i,1}(\mathbf{s}_{-i}) &= \mathbf{K}_{i,0}(\mathbf{s}_{-i}) \;+\; \mathbf{X}_i \mathbf{R}_i \mathbf{X}_i^{\sf H}.
\end{align}

\subsubsection{Design Objective and FP-KLD Instantiation}
Define the joint distributions over the latent interference state and the snapshots:
\begin{align}
    P_{i,0}(\mathbf{s}_{-i},\mathbf{Y}) &= w(\mathbf{s}_{-i})\,p_\mathcal{CN}\big(\bar{\mathbf{y}};\, \mathbf{I}_{N_r}\otimes\mathbf{K}_{i,0}(\mathbf{s}_{-i})\big),\\
    P_{i,1}(\mathbf{s}_{-i},\mathbf{Y}) &= w(\mathbf{s}_{-i})\,p_\mathcal{CN}\big(\bar{\mathbf{y}};\, \mathbf{I}_{N_r}\otimes\mathbf{K}_{i,1}(\mathbf{s}_{-i})\big),
\end{align}
where $\bar{\mathbf{y}}= \operatorname{vec}(\mathbf{Y})$. Since the prior $w(\mathbf{s}_{-i})$ is common to both hypotheses, the chain rule gives
\begin{align}
    &D_{\mathrm{KL}}\big(P_{i,0}(\mathbf{s}_{-i},\mathbf{Y})\,\big\|\,P_{i,1}(\mathbf{s}_{-i},\mathbf{Y})\big)\nonumber\\
    &= \mathbb{E}_{\mathbf{s}_{-i}\sim w}\Big[
       D_{\mathrm{KL}}\big(p_\mathcal{CN}(\bar{\mathbf{y}}; \mathbf{I}_{N_r}\otimes\mathbf{K}_{i,0}(\mathbf{s}_{-i})) \nonumber\\
    &\qquad\qquad\qquad \big\| \,
       p_\mathcal{CN}(\bar{\mathbf{y}}; \mathbf{I}_{N_r}\otimes\mathbf{K}_{i,1}(\mathbf{s}_{-i}))\big)
    \Big].
\end{align}
Therefore, the joint error exponent for user $i$ equals a weighted sum of Gaussian KLDs. Summing over users yields the multi-user design objective to be maximized as:
\begin{align}
    D_{\Sigma}(\{\mathbf{X}_i\})
    &\triangleq \sum_{i=1}^K D_{\mathrm{KL}}\big(P_{i,0}\Vert P_{i,1}\big)\nonumber\\
    &= N_r \sum_{i=1}^K \sum_{\mathbf{s}_{-i}\in\CMcal{S}_{-i}} w(\mathbf{s}_{-i})
       \Big(\log \big|\mathbf{K}_{i,0}^{-1}(\mathbf{s}_{-i})\mathbf{K}_{i,1}(\mathbf{s}_{-i})\big|
            \nonumber\\
            &+ \operatorname{Tr}(\mathbf{K}_{i,1}^{-1}(\mathbf{s}_{-i})\,\mathbf{K}_{i,0}(\mathbf{s}_{-i})) - T \Big).
    \label{eq:obj_randomaccess}
\end{align}
We maximize \eqref{eq:obj_randomaccess} subject to per-device power constraints $\operatorname{Tr}(\mathbf{X}_i\mathbf{X}_i^{\sf H})\leq P_t$. Finally, the optimization problem formulation follows as:
\begin{align} \label{eq:random_access_design}
    \underset{\{\mathbf{X}_i\}_{i=1}^K}{\text{maximize}}\quad &D_\Sigma (\{\mathbf{X}_i\})\\
    \text{subject to}\quad & \operatorname{Tr}(\mathbf{X}_i\mathbf{X}_i^{\sf H}) \leq P_t, \quad \text{for}\ i=1,\cdots,K. \nonumber
\end{align}
Since each term in \eqref{eq:random_access_design} has the same algebraic form as in the single-Gaussian case, our algorithms introduced in previous sections (e.g., FP-KLD and A-MM-KLD) are suitably applied. 
In the next section, we show that the waveforms optimized via our A-MM-KLD significantly outperform traditional designs in terms of detection probability.

\section{Numerical Results}\label{sec:results}
In this section, we evaluate the performance of the proposed FP-based optimization methods via numerical simulations. We validate the effectiveness of our approach in three representative scenarios: 1) MIMO radar waveform design for binary hypothesis testing, 2) joint ISAC waveform design, and 3) active user detection in multiple access systems. The convergence tolerance for all iterative algorithms is set to $\tau=10^{-6}$.

\subsection{Convergence and Runtime Analysis}

\begin{figure}
    \centering
    \includegraphics[width=0.75\linewidth]{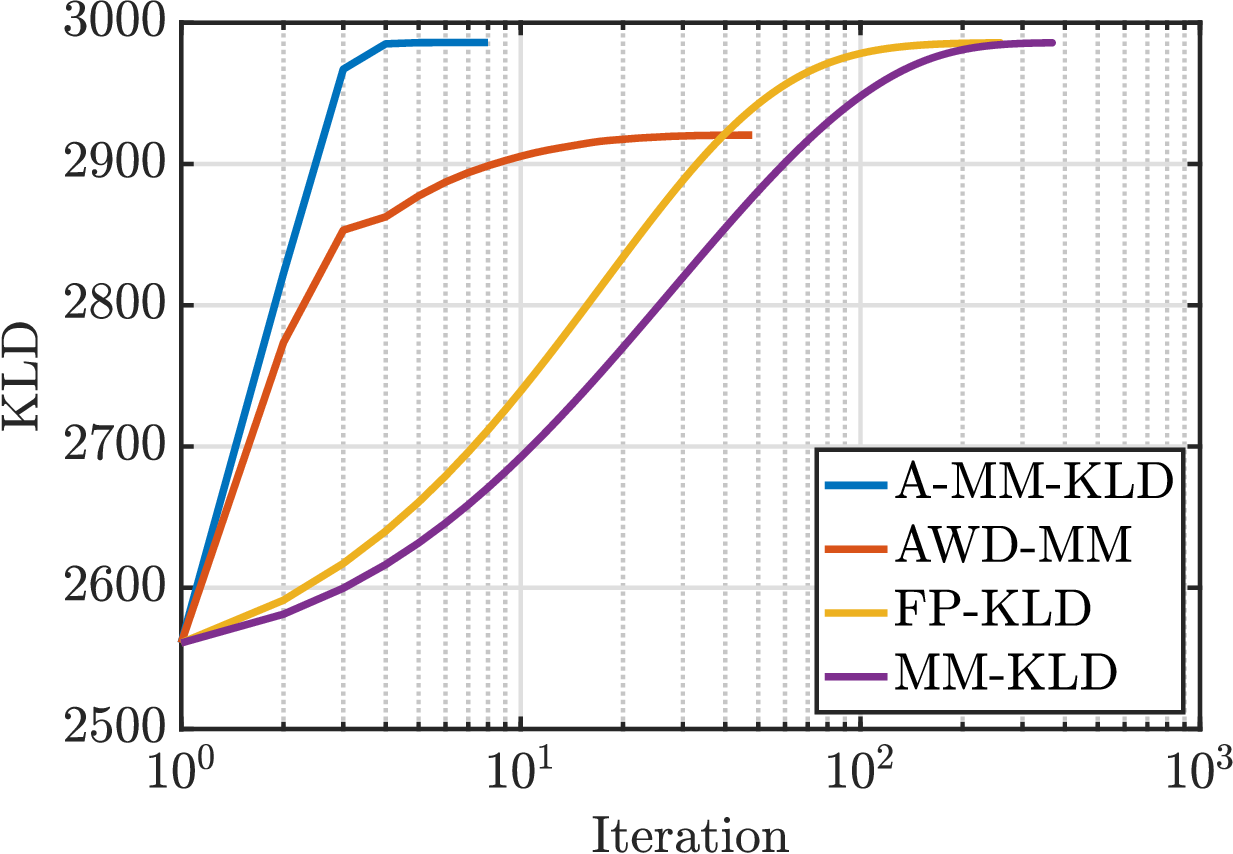}
    \caption{Convergence profile of the objective function versus the number of iterations. The proposed algorithms are compared with {AWD-MM}.}
    \label{fig:KLDCurves}
\end{figure}

We first assess the convergence behavior and computational efficiency of the proposed algorithms. For the KLD maximization, the covariance matrices $\bfR_H, \bfR_0$, and $\bfR_1$ are independently generated for each random environment. Each covariance is constructed as $\bfU\mathbf{\Lambda}\bfU^\herm$, where $\bfU$ is obtained from the QR decomposition of an i.i.d. complex Gaussian random matrix, yielding an isotropically random unitary eigenbasis. The eigenvalues of $\bfR_H$ are drawn i.i.d. from $\rm{Unif}[1,5]$, while those of $\bfR_0$ and $\bfR_1$ are drawn i.i.d. from $\rm{Unif}[0.1, 0.5]$. We consider a MIMO sensing setup with $N_t=N_r=32$ antennas and a sequence length of $T=50$. The SNR is set to 7 dB. We compare our proposed frameworks--FP-KLD (Algorithm~\ref{alg:FP-KLD}), MM-KLD (Algorithm~\ref{alg:MM-KLD}), and A-MM-KLD (Algorithm~\ref{alg:A-MM-KLD})--against the state-of-the-art benchmark, AWD-MM.

\begin{figure}[t]
    \centering
    \subfloat[Runtime per iteration\label{fig:runtime_per_iter}]{\includegraphics[width=0.78\columnwidth]{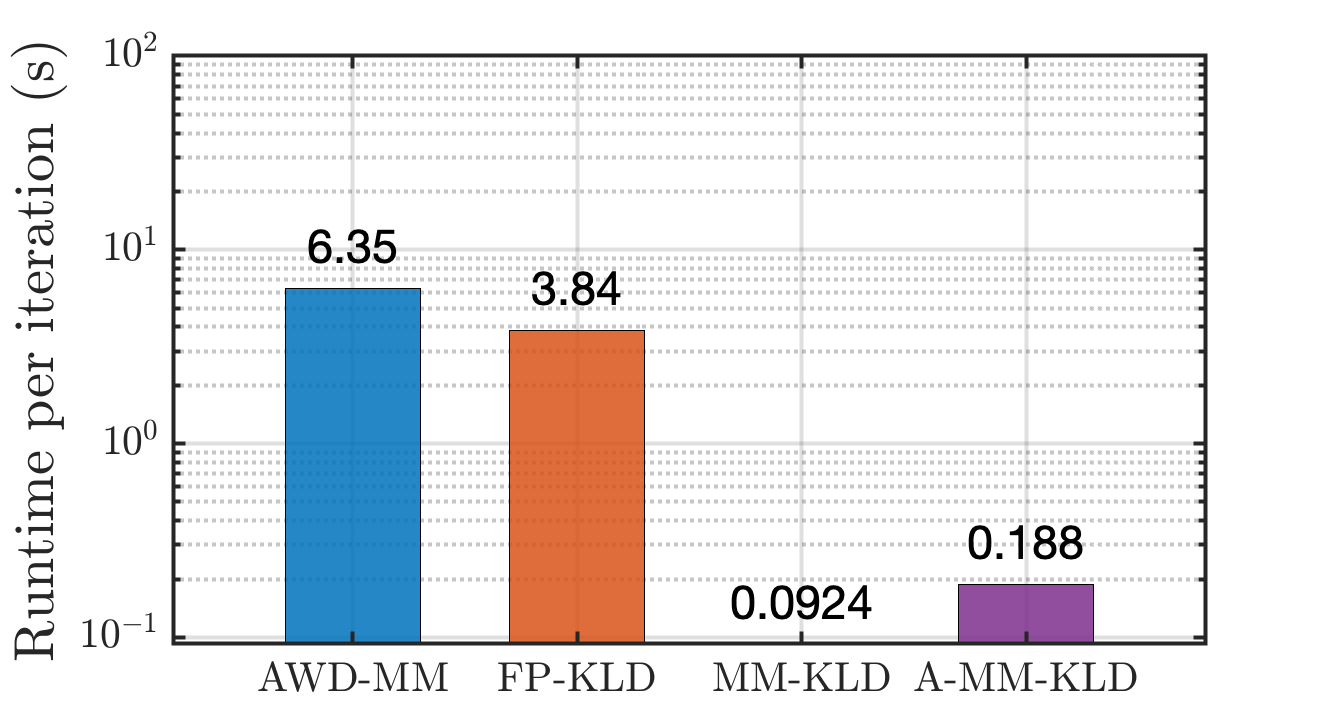}}\vfill
    \subfloat[Total runtime\label{fig:runtime_total}]{\includegraphics[width=0.78\columnwidth]{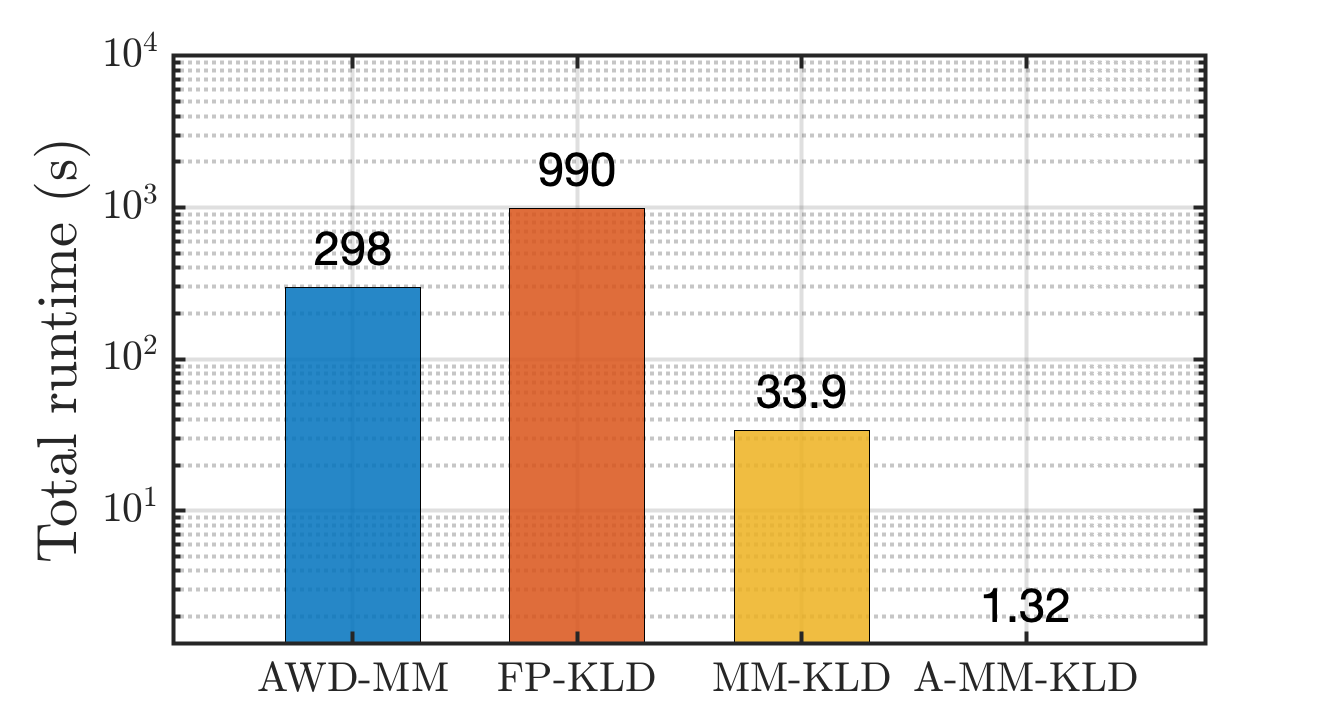}}
    \caption{Computational complexity comparison in terms of runtime. The proposed {A-MM-KLD} algorithm achieves orders-of-magnitude lower runtime than the others.}
    \label{fig:runtime}
\end{figure}

Fig.~\ref{fig:KLDCurves} illustrates the evolution of the KLD objective function with respect to the number of iterations. Several key observations can be made regarding the convergence properties of the proposed schemes. First, all proposed algorithms exhibit a monotonic increase in the objective function, empirically validating our theoretical derivation that frames them as valid instances of the MM framework. Second, comparing the unaccelerated schemes, FP-KLD requires fewer iterations to converge than MM-KLD. This behavior directly corroborates our theoretical analysis in Section~\ref{subsec:relaxation}: FP-KLD utilizes a tight, anisotropic quadratic surrogate derived from the matrix fractional programming transform, whereas MM-KLD employs a nonhomogeneous relaxation. This relaxation based on Lemma~\ref{lem:nonhomogeneous} replaces the exact curvature with a conservative isotropic spectral bound, resulting in a looser surrogate and, consequently, a slower linear convergence rate.
However, the proposed A-MM-KLD dramatically overcomes this limitation. By incorporating STEM, it effectively uses secant-type information of the fixed-point mapping and exhibits much faster empirical convergence. As shown in Fig.~\ref{fig:KLDCurves}, A-MM-KLD converges rapidly requiring fewer iterations than the computationally heavier methods.

While iteration count is an important metric, the practical latency of an algorithm is determined by the total runtime, which is the product of the number of iterations and the per-iteration computation time. Fig.~\ref{fig:runtime} presents a comprehensive runtime analysis.

Fig.~\ref{fig:runtime_per_iter} compares the average runtime required for a single iteration. Both AWD-MM and FP-KLD suffer from high per-iteration cost because they involve solving large-scale linear systems at every step. In contrast, MM-KLD and A-MM-KLD reduce the per-iteration complexity by replacing the linear-system update with a closed-form waveform update.

Fig.~\ref{fig:runtime_total} displays the total runtime until convergence. The results highlight the superiority of the proposed A-MM-KLD framework. Compared with AWD-MM, A-MM-KLD retains the same KLD-based detection objective but achieves substantially lower runtime by combining the nonhomogeneous relaxation with acceleration.

\subsection{Detection performance under Neyman--Pearson Criterion}
\begin{figure}[!t]
    \centering
    \includegraphics[width=0.7\linewidth]{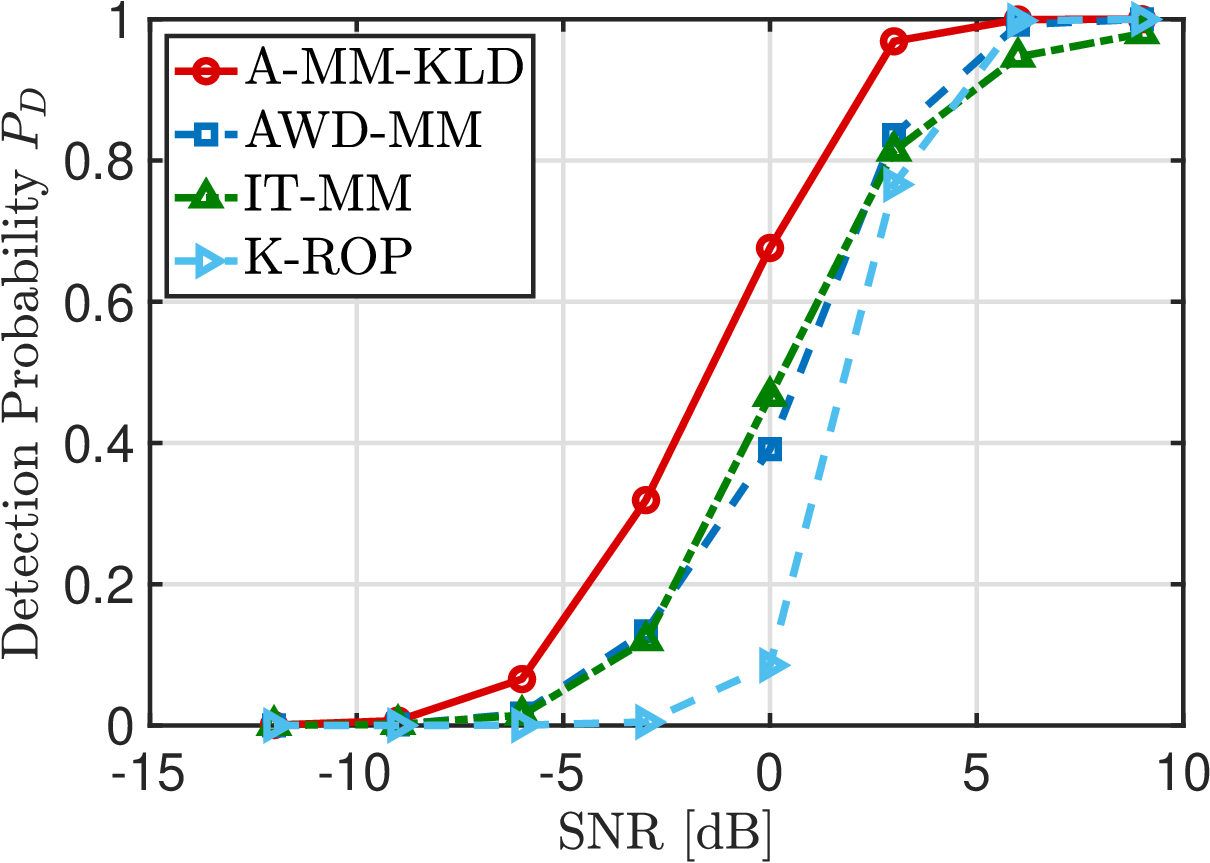}
    \caption{Detection probability $P_D$ versus SNR under a fixed false-alarm probability.}
    \label{fig:baseline_detection_pd}
\end{figure}

While the previous results focus on convergence and runtime, we further evaluate the detection performance of the optimized waveforms under the Neyman--Pearson criterion. To make the comparison more comprehensive, we compare the proposed A-MM-KLD with three optimization-based sensing waveform-design baselines: AWD-MM, IT-MM, and K-ROP. IT-MM is an information-theoretic MM radar code-design baseline in \cite{naghsh2013unified}, while K-ROP is a KLD-based projected-gradient radar waveform-design baseline in \cite{baseline1}. All baseline methods are used only to generate transmit waveforms, and the final detection probability is evaluated using the same Neyman--Pearson LRT under the proposed cluttered Gaussian sensing model.

In this experiment, we use $N_t=N_r=8$ transmit/receive antennas and waveform length $T=16$. For each SNR point, the results are averaged over $N_{\rm cov}=10$ independent covariance environments. The false-alarm probability is fixed as $\alpha=10^{-5}$. The LRT threshold is obtained from $N_\eta=10^6$ Monte Carlo samples under $\CMcal{H}_0$, and the detection probability is estimated using $N_{\rm PD}=2\times 10^5$ Monte Carlo samples under $\CMcal{H}_1$.

Fig.~\ref{fig:baseline_detection_pd} shows the detection probability versus SNR. The proposed A-MM-KLD achieves the fastest transition to high detection probability among the baselines. AWD-MM and IT-MM require higher SNR to reach comparable detection probability, whereas K-ROP approaches high detection probability only at sufficiently high SNR. This confirms that the proposed method improves not only optimization efficiency but also the final detection performance in the considered hypothesis-testing task.

\subsection{Initialization Sensitivity}
\label{subsec:init_sensitivity}

\begin{figure}[t]
  \centering
  \subfloat[Final KLD distributions\label{fig:init_alg_a}]{\includegraphics[width=0.9\columnwidth]{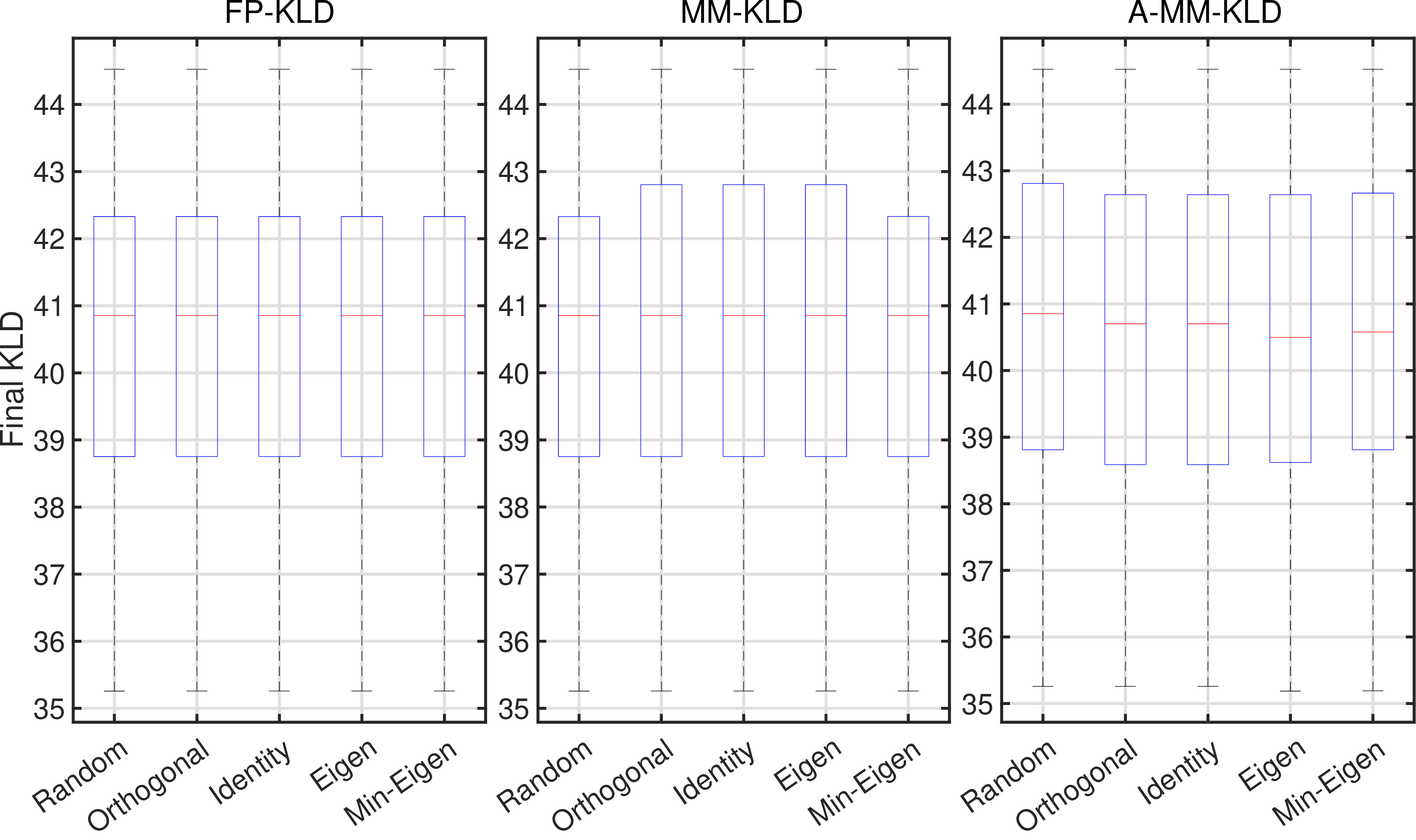}}\\
  \subfloat[Iteration-count distributions\label{fig:init_alg_b}]{\includegraphics[width=0.9\columnwidth]{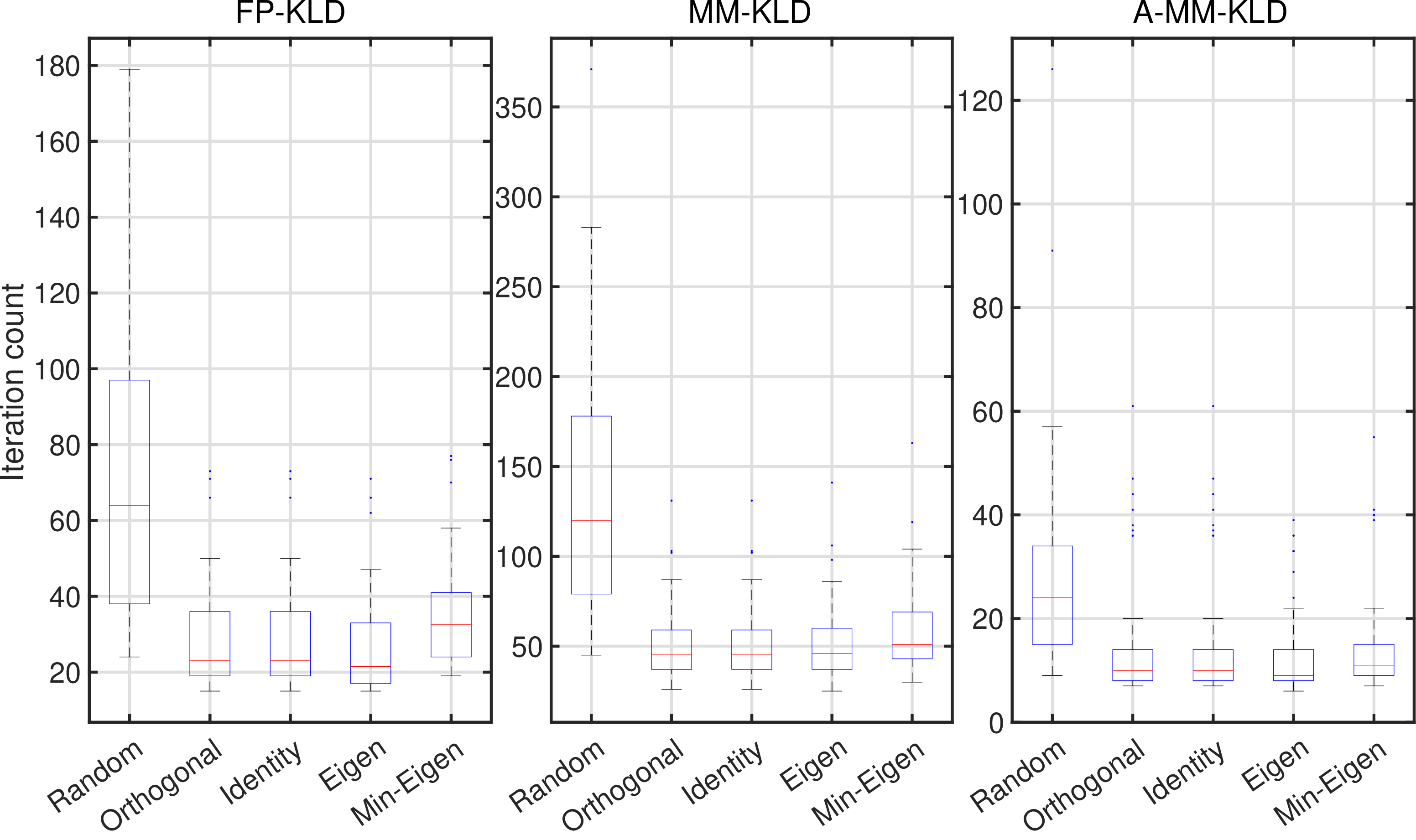}}
  \caption{Initialization sensitivity of FP-KLD, MM-KLD, and A-MM-KLD}
  \label{fig:init_alg}
\end{figure}

Since the proposed algorithms are MM-type methods for a nonconvex problem, their solution quality and iteration count may in principle depend on the initialization. We evaluate five initializations of the transmit waveform $\mathbf{X}^{(0)}$, all normalized to satisfy the transmit-power constraint. The random Gaussian initialization draws $\mathbf{X}^{(0)}$ with i.i.d. circularly symmetric complex Gaussian entries. The QR-based orthogonal initialization uses the thin-QR factor of an i.i.d. complex Gaussian matrix. The identity-like initialization uses $\mathbf{X}^{(0)}=[\mathbf{I}_{N_t} \mathbf{0}_{N_t\times (T-N_t)}]^\trans$ before normalization. The eigen-aligned initialization uses the dominant eigendirections of $\mathbf{R}_{H1}-\mathbf{R}_0$ as a covariance-aware warm start. The minimum-eigenvalue initialization uses the eigendirections corresponding to the smallest eigenvalues of $\mathbf{R}_{H1}-\mathbf{R}_0$ and is included as a low-informativeness stress test. We use $N_t=N_r=16$, $T=30$, and 50 independent covariance environments.

\begin{figure}[t]
  \centering
  \subfloat[Final KLD distributions\label{fig:init_snr_a}]{\includegraphics[width=0.85\columnwidth]{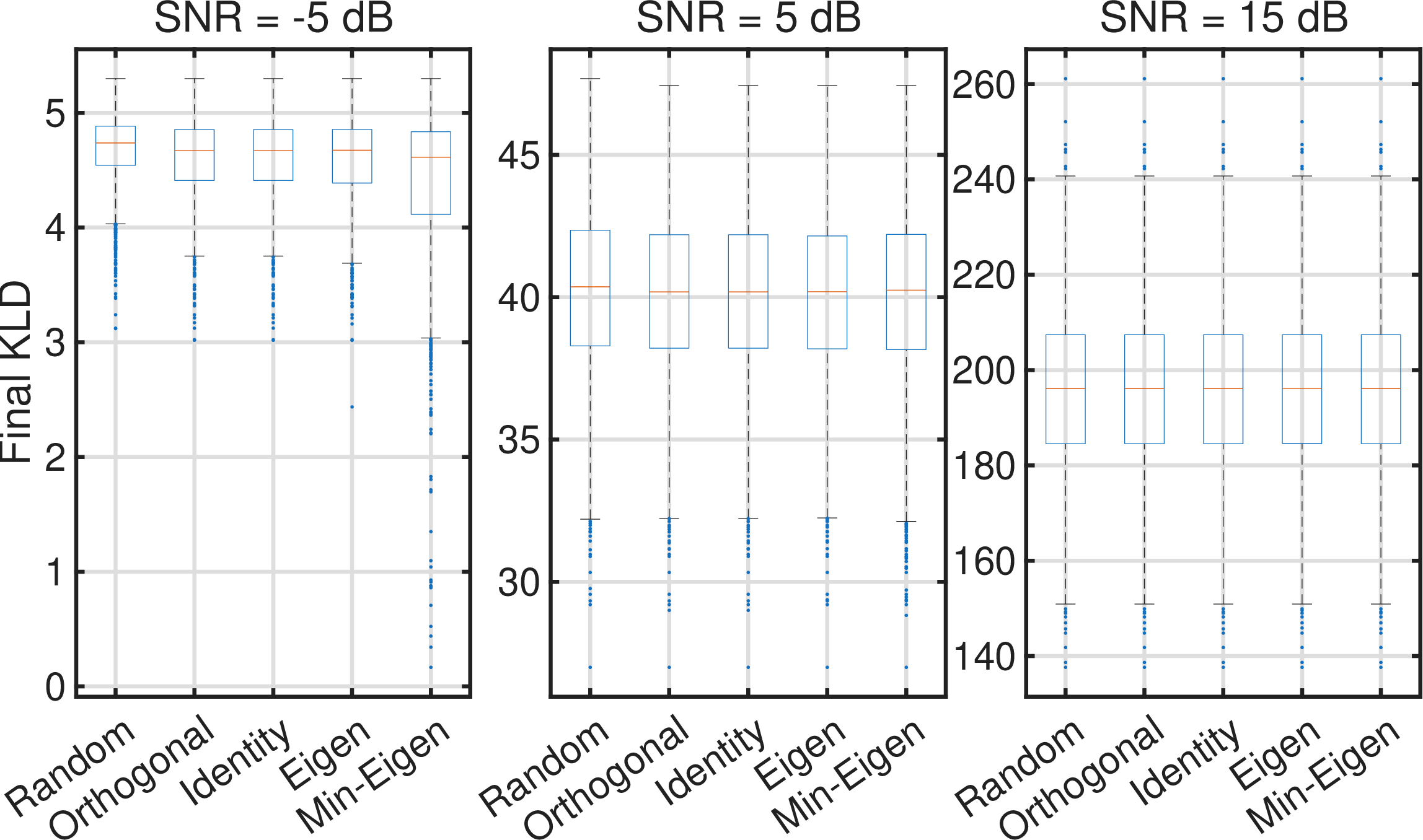}}\\
  \subfloat[Iteration count\label{fig:init_snr_b}]{\includegraphics[width=0.75\columnwidth]{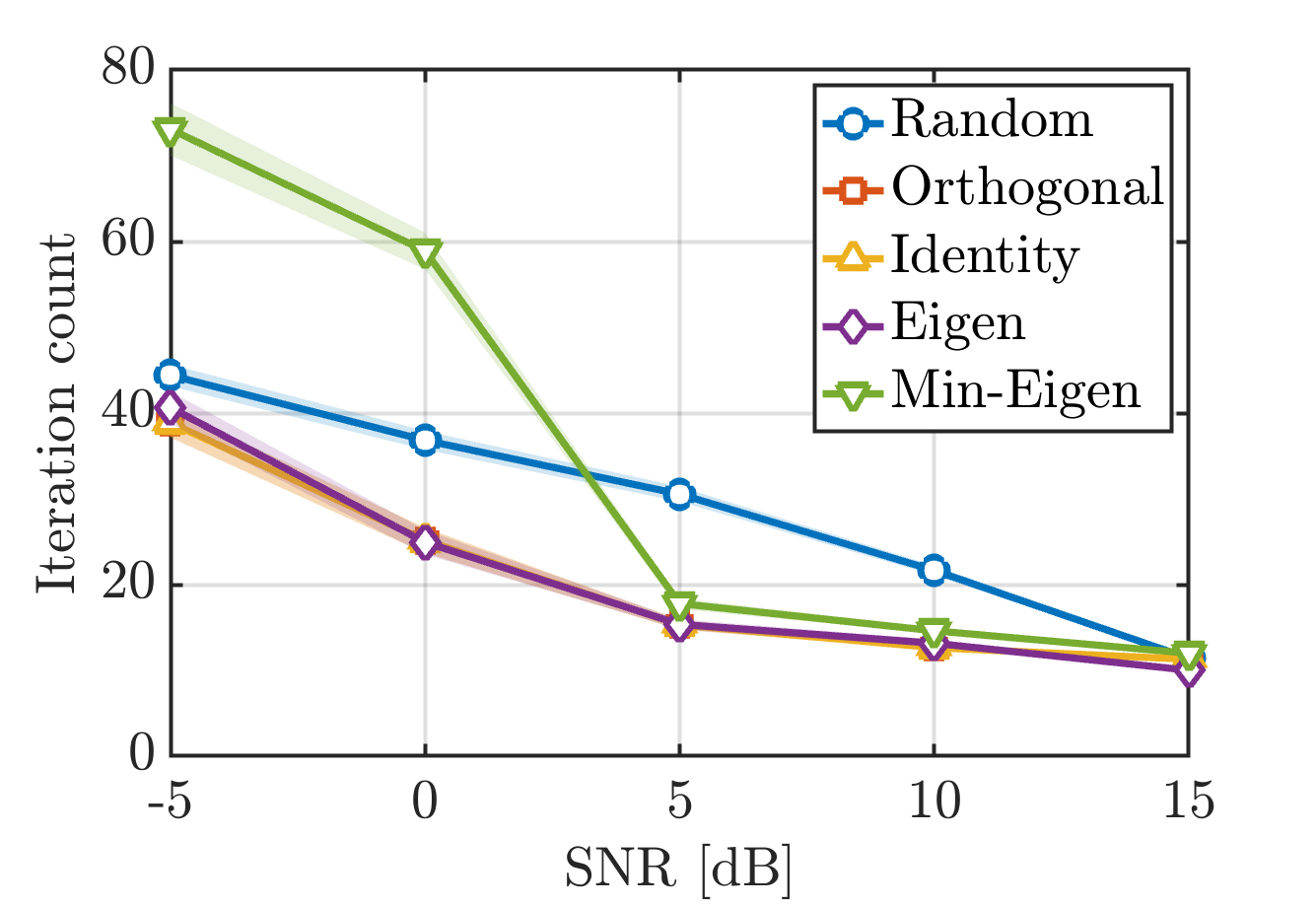}}
  \caption{Initialization sensitivity of A-MM-KLD versus SNR}
  \label{fig:init_snr}
\end{figure}

Fig.~\ref{fig:init_alg} compares the three algorithms at $\mathrm{SNR}=5$~dB. In Fig.~\ref{fig:init_alg_a} the final KLD distributions overlap closely across all initializations and algorithms, so the attained solution quality is essentially insensitive to both. The iteration count in Fig.~\ref{fig:init_alg_b} is more sensitive: structurally favorable starts reduce it relative to a random start, and A-MM-KLD requires the fewest iterations among the three algorithms, consistent with its acceleration. As shown in Fig.~\ref{fig:init_snr_a}, the robustness of the final KLD to the initialization is maintained across SNR, while Fig.~\ref{fig:init_snr_b} shows that the reduction in iteration count from the structurally favorable starts also persists.

These trends can be explained by the conditioning and covariance alignment of the initial waveform. A random Gaussian start is normalized to the same total power but can still have uneven column powers and non-negligible inter-column correlations, whereas the orthogonal and identity-like starts are well-conditioned and power-balanced. The eigen-aligned start further places energy in the dominant eigendirections of $\mathbf{R}_{H1}-\mathbf{R}_0$, while the minimum-eigenvalue start is intentionally unfavorable. Nevertheless, the MM-type updates can reshape the waveform toward informative directions, which explains why the final KLD is robust while the iteration count is more sensitive. Accordingly, an orthogonal or identity-like initialization is recommended as a robust default, and the eigen-aligned initialization as a principled warm start when reliable covariance statistics are available.

\subsection{Effect of Covariance Mismatch}

Because the waveform is designed from long-term covariance statistics, a practically relevant question is how the design degrades when the assumed target covariance differs from the true one. To examine this effect, the transmitter designs the waveform using a nominal target covariance $\bfR_{H,{\rm nom}}$, while the performance is evaluated under a mismatched true covariance
\begin{align}
    \bfR_{H,{\rm true}}
    =
    (1-\epsilon)\bfR_{H,{\rm nom}}
    +
    \epsilon\bfR_{H,{\rm hidden}},
    \label{eq:cov_mismatch}
\end{align}
where $\epsilon\in[0,1]$ denotes the fraction of an unmodeled hidden covariance component. The total target covariance power is preserved for all $\epsilon$, so that the observed degradation is due to covariance-structure mismatch rather than a change in target power. We sweep $\epsilon$ and report the relative KLD normalized by the full-knowledge oracle, which designs the waveform using $\bfR_{H,{\rm true}}$.
\begin{figure}[t]
    \centering
    \includegraphics[width=0.7\columnwidth]{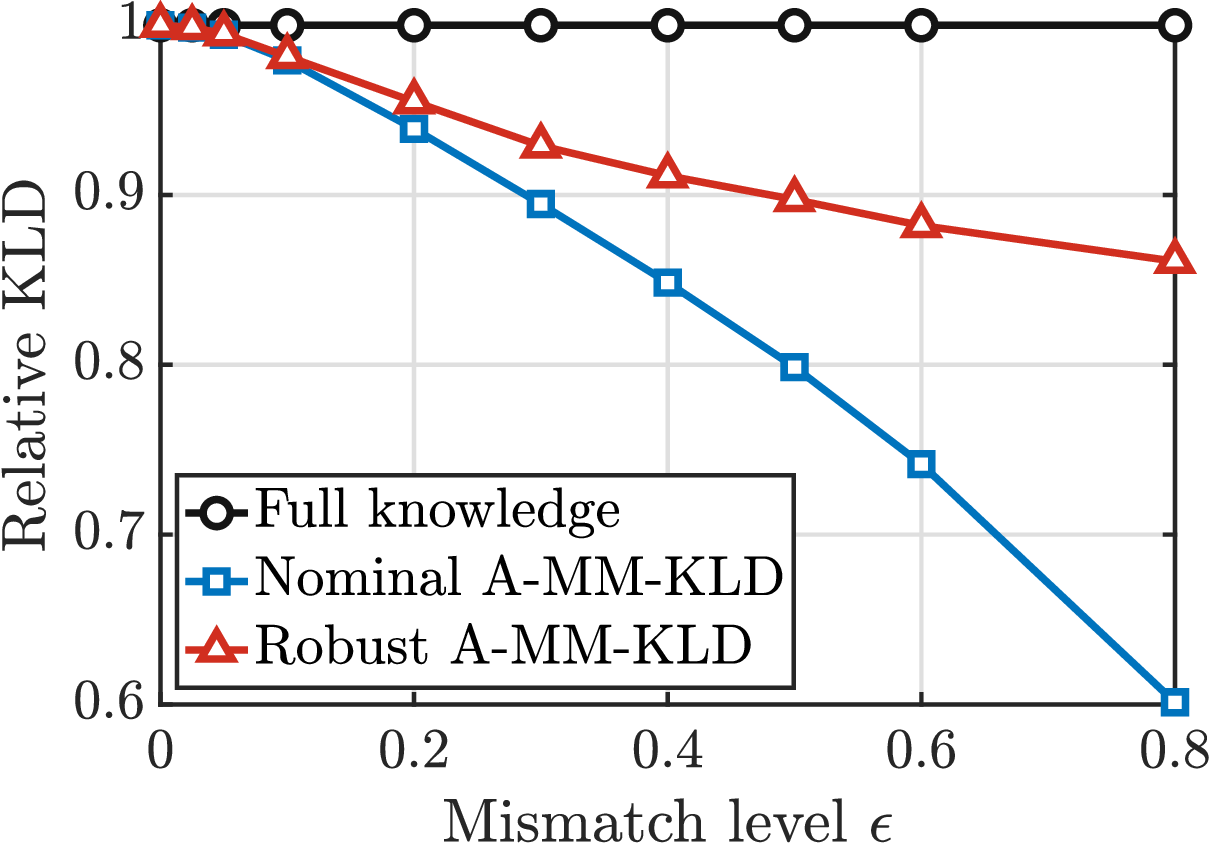}
    \caption{Relative KLD versus the covariance mismatch level $\epsilon$, normalized by the full-knowledge oracle reference.}
    \label{fig:covariance_mismatch}
\end{figure}

The nominal A-MM-KLD design solves the original KLD maximization problem using only the assumed covariance,
\begin{align}
    \underset{\bfX}{\text{maximize}}\ 
    D_{\rm KL}(\bfX;\bfR_{H,{\rm nom}})
    \quad
    \text{s.t.}\ \|\bfX\|_F^2\le P_t .
\end{align}
As shown in Fig.~\ref{fig:covariance_mismatch}, this design closely tracks the full-knowledge oracle under mild mismatch, but becomes increasingly sensitive as $\epsilon$ grows.

To mitigate this sensitivity, we consider a leakage-aware variant that modifies only the covariance used in the design objective. Since the actual hidden component is unknown to the transmitter, the design uses a finite ensemble of plausible hidden covariances. Specifically, let $\bfV_{\rm low}$ denote the weak eigenspace of $\bfR_{H,{\rm nom}}$, formed by the eigenvectors associated with its smallest eigenvalues. For representative unit-norm directions $\bfv_m\in{\rm span}(\bfV_{\rm low})$, we construct
\begin{align}
    \bfR_{H,{\rm hidden},m}
    =
    \operatorname{Tr}(\bfR_{H,{\rm nom}})
    \bfv_m\bfv_m^{\sf H},
    \qquad
    \|\bfv_m\|_2=1,
\end{align}
and form the candidate covariances
\begin{align}
    \bfR_m(\epsilon)
    =
    (1-\epsilon)\bfR_{H,{\rm nom}}
    +
    \epsilon\bfR_{H,{\rm hidden},m}.
\end{align}
The leakage-aware waveform is then obtained by maximizing the averaged KLD
\begin{align}
    \underset{\bfX}{\text{maximize}}\ 
    \frac{1}{M}\sum_{m=1}^{M}
    D_{\mathrm{KL}}(\bfX;\bfR_m(\epsilon))
    \quad
    \text{s.t.}\ \|\bfX\|_{F}^{2}\le P_t .
    \label{eq:leakage}
\end{align}
This design does not require knowledge of the specific hidden direction used in the evaluation; it only uses representative leakage patterns. Since the objective remains a sum of single-Gaussian KLD terms, it is algebraically analogous to the multiple random-access extension in Section~\ref{subsec:random_access}, and the FP/MM surrogate construction applies scenario-wise. As shown in Fig.~\ref{fig:covariance_mismatch}, the robust design remains much closer to the full-knowledge oracle, recovering a substantial portion of the loss incurred by the nominal design. Thus, the proposed framework can reduce mismatch sensitivity within the same KLD maximization formulation by replacing the single nominal covariance with an uncertainty ensemble.

\subsection{ISAC Rate--Detection Trade-off}\label{subsec:isac_results}
\begin{figure}[t]
  \centering
  \includegraphics[width=0.68\columnwidth]{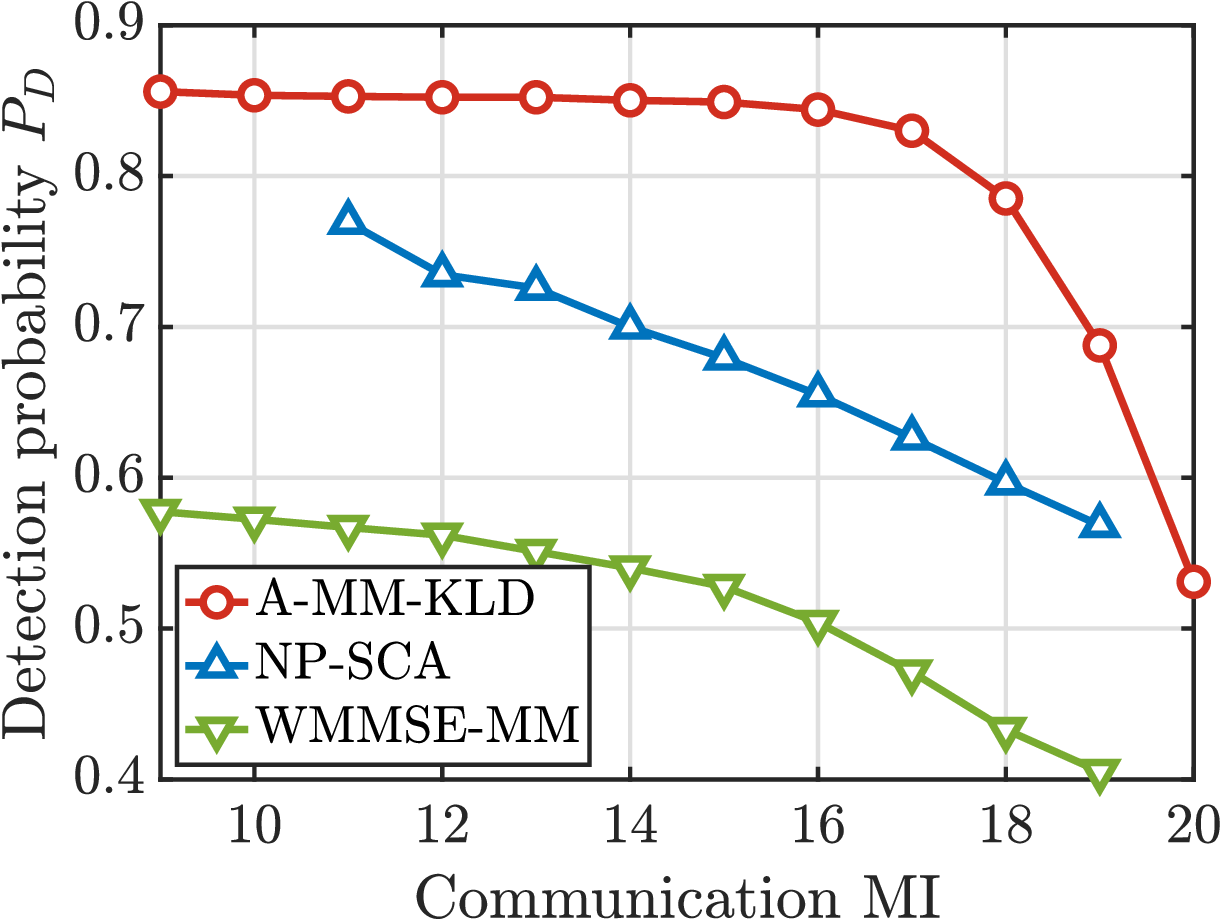}
  \caption{ISAC rate--detection comparison under common communication MI and Neyman--Pearson detection metrics.}
  \label{fig:isac_baseline}
\end{figure}

 We now evaluate the ISAC waveform design of Section~\ref{subsec:isac} by comparing it with representative established ISAC waveform-design methods under common communication and sensing metrics. Specifically, we compare the proposed A-MM-KLD design with a successive convex approximation (SCA)-based ISAC design (NP-SCA) \cite{song2025detection} and a weighted minimum mean square error (WMMSE)/MM-based ISAC design (WMMSE-MM) \cite{xu2020multi}. For a fair comparison, all generated waveforms are evaluated using the same metrics: the communication performance is measured by the MI in \eqref{eq:comm_mi}, and the sensing performance is measured by the detection probability of the Neyman--Pearson LRT in \eqref{eq:lrt} under the same fixed false-alarm probability.

In this experiment, we use $N_t=N_r=10$, $K=10$, $T=12$, sensing/communication SNRs of $0/10$~dB, and fixed false-alarm probability $\alpha=10^{-5}$. The results are averaged over $N_{\rm env}=100$ independent environments, with the LRT threshold and detection probability estimated using $2\times 10^5$ samples drawn under $\CMcal{H}_0$ and $\CMcal{H}_1$, respectively. Fig.~\ref{fig:isac_baseline} shows that the proposed design achieves higher detection probability than NP-SCA and WMMSE-MM over the considered communication rate range. This result demonstrates that the proposed design framework can be effectively applied to the considered ISAC objective and provides a more favorable rate--detection trade-off than the baseline methods.

\subsection{Performance in Multiple Random Access}

We next apply the proposed A-MM-KLD algorithm to the multiple random access scenario described in Section~\ref{subsec:random_access}. We consider a system with $K=4$ devices, where both the transmitter and receiver are equipped with $N_t=N_r=4$ antennas. The per-user covariance is generated as $\bfR_{H,k}=\mathbf{A}\mathbf{A}^\herm$ followed by trace normalization, where $\mathbf{A}$ has i.i.d. complex Gaussian entries. For detection, we employ the LRT based on the Neyman--Pearson criterion. The detection threshold $\eta$ is determined numerically to satisfy a target probability of false alarm, set to $\alpha = 10^{-3}$. The performance metric is the geometric mean of the detection probabilities of all users, which ensures fairness and reflects the overall system reliability. We compare our optimized waveforms against conventional orthogonal sequences.

\begin{figure}[!t]
    \centering
    \subfloat[Detection probability versus SNR ($T=8$).\label{fig:random_access_snr}]{
        \includegraphics[width=0.46\linewidth]{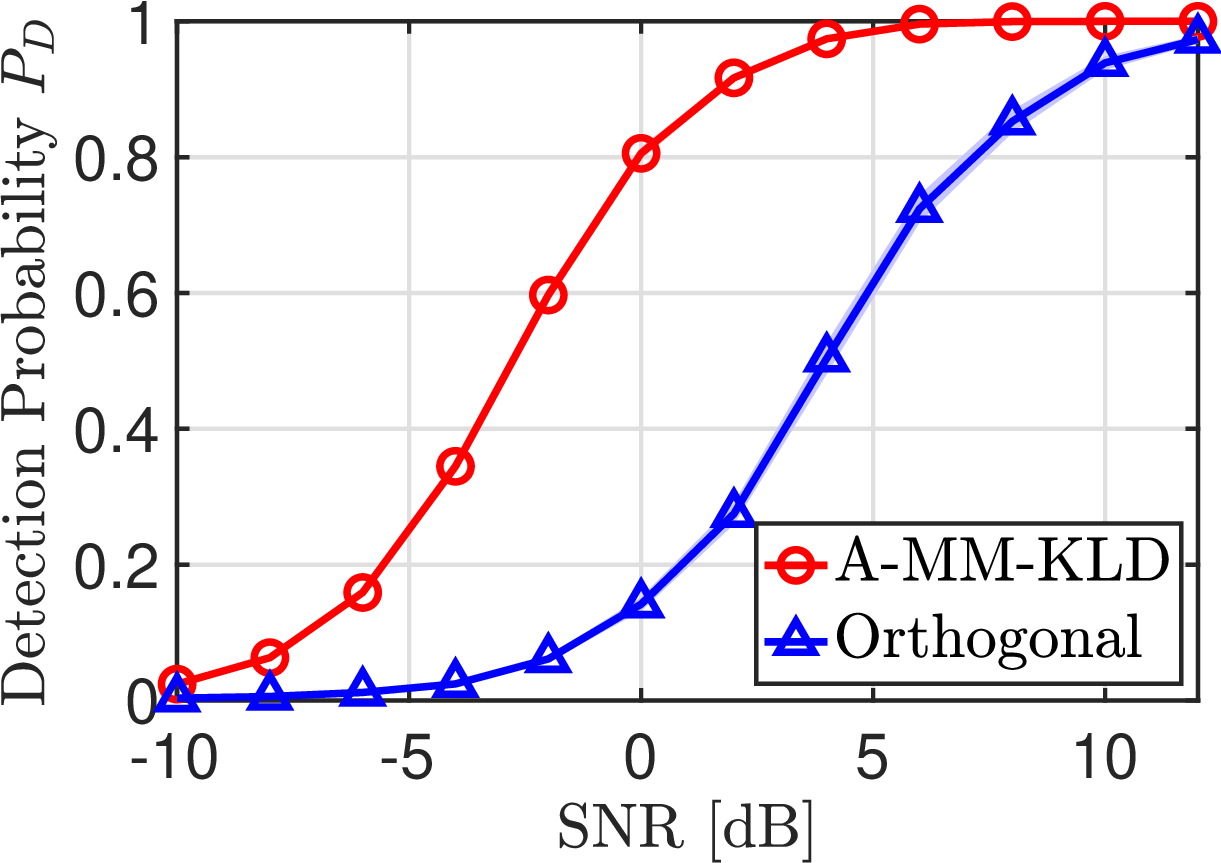}}
    \hfill
    \subfloat[Detection probability versus sequence length ($\mathrm{SNR}=8$~dB).\label{fig:random_access_T}]{
        \includegraphics[width=0.46\linewidth]{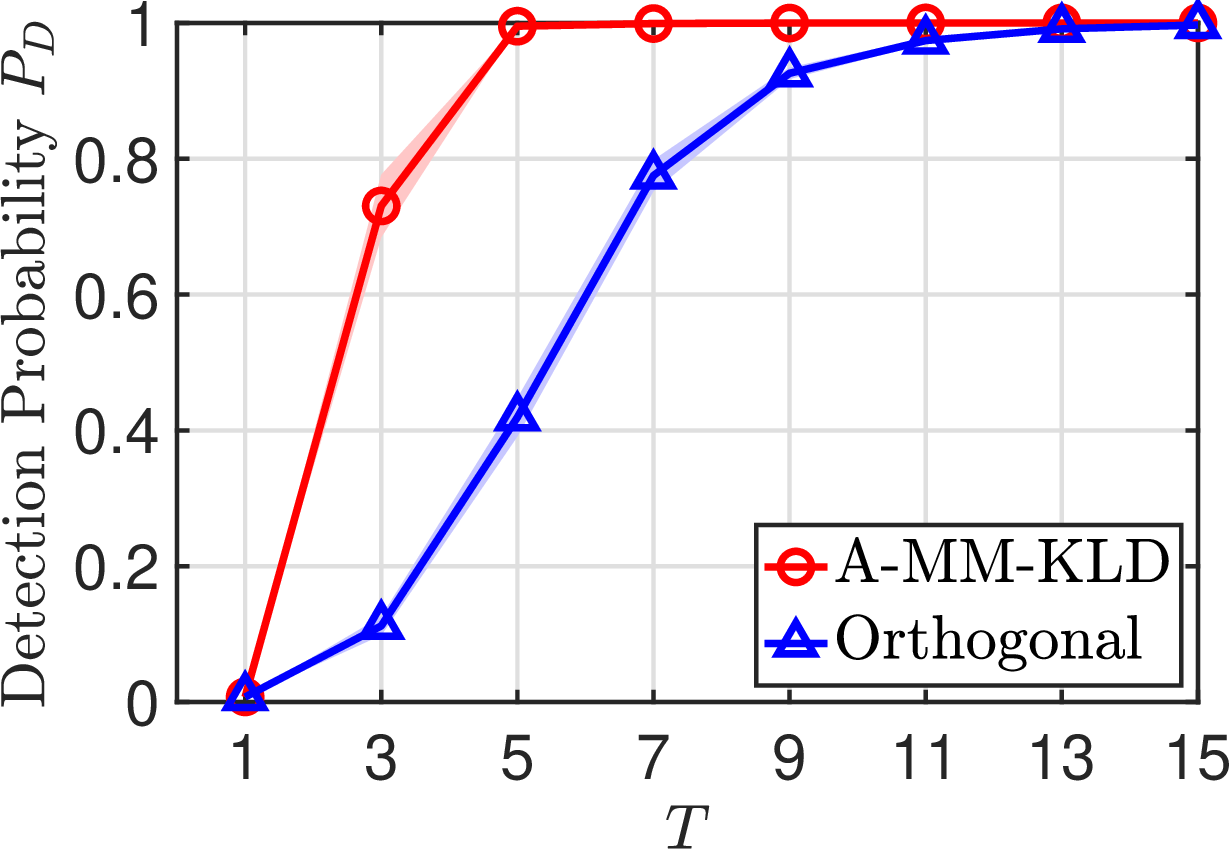}}
    \caption{Geometric mean of detection probabilities of the proposed A-MM-KLD design and orthogonal sequences.}
    \label{fig:random_access}
\end{figure}

The detection performance is evaluated via Monte Carlo simulation. For each random environment generated according to the above setup, the LRT threshold $\eta$ is estimated using $N_\eta=10^5$ samples drawn under $\CMcal{H}_0$, and the per-user detection probability under $\CMcal{H}_1$ is then estimated using $N_{MC}=10^5$ samples. The reported curves are obtained by averaging the per-environment geometric-mean detection probability over $N_{env}=100$ independent random environments. To quantify across random environments, including Monte Carlo estimation effects, Fig.~\ref{fig:random_access} additionally includes shaded $95\%$ confidence bands computed as $\bar{P}_D\pm1.96 \hat{\sigma}/\sqrt{N_{env}}$, where $\bar{P}_D$ and $\hat{\sigma}$ denote the sample mean and standard deviation of the environment-level geometric-mean detection probabilities. The confidence bands are sufficiently narrow that they largely overlap with the mean curves and are substantially smaller than the performance gap between the proposed and baseline schemes, confirming that the reported gains are statistically significant.

Fig.~\ref{fig:random_access_snr} presents the detection performance versus SNR with a fixed sequence length of $T=8$. It is observed that the proposed A-MM-KLD algorithm yields a substantial performance gain over the orthogonal sequences across the entire SNR regime. This advantage stems from the fact that fixed orthogonal sequences do not account for the specific spatial covariance structures of the channels or the interference patterns. In contrast, our method explicitly maximizes the weighted sum of KLDs by jointly optimizing the waveforms to be robust against MUI generated by the random activity of other devices. Consequently, the proposed design achieves reliable detection even at lower SNR levels where the baseline fails.

Fig.~\ref{fig:random_access_T} illustrates the detection performance versus the sequence length $T$ at a fixed SNR of $8$ dB. As expected, the detection probability improves as $T$ increases for both schemes, since a longer duration provides more degrees of freedom to suppress interference and accumulate signal energy. However, the proposed algorithm demonstrates significantly faster saturation to perfect detection. Notably, in the regime of short sequence lengths, where the system is overloaded, the orthogonal sequences suffer severely from MUI. Conversely, the proposed waveforms maintain high detection capability even with limited temporal resources, highlighting the spectral efficiency and low-latency potential of our KLD-based design.

So far, all numerical experiments have relied on synthetically generated covariance matrices. To further demonstrate that the observed gains are not artifacts of this particular statistical model, we additionally validate the random-access design using site-specific full-MIMO channels generated by the DeepMIMO ray-tracing framework~\cite{alkhateeb2019deepmimo}. Each DeepMIMO scenario is obtained by ray tracing over a specific three-dimensional environment and encodes physical scene descriptors---including the scene geometry, the electromagnetic/material properties of surfaces, and the modeled propagation interactions such as reflection, diffraction, scattering, and transmission. The resulting channel realizations are thus site-specific, geometry-consistent, and fully reproducible from the published scenario and parameter set. While DeepMIMO does not constitute over-the-air measured channel state information, it yields realistic, environment-dependent spatial covariance statistics, providing a complementary and physically grounded test of the proposed waveform design.

\begin{figure}[!t]
    \centering
    \subfloat[Detection probability versus SNR ($T=8$).\label{fig:rt_ra_prob_snr}]{
        \includegraphics[width=0.46\linewidth]{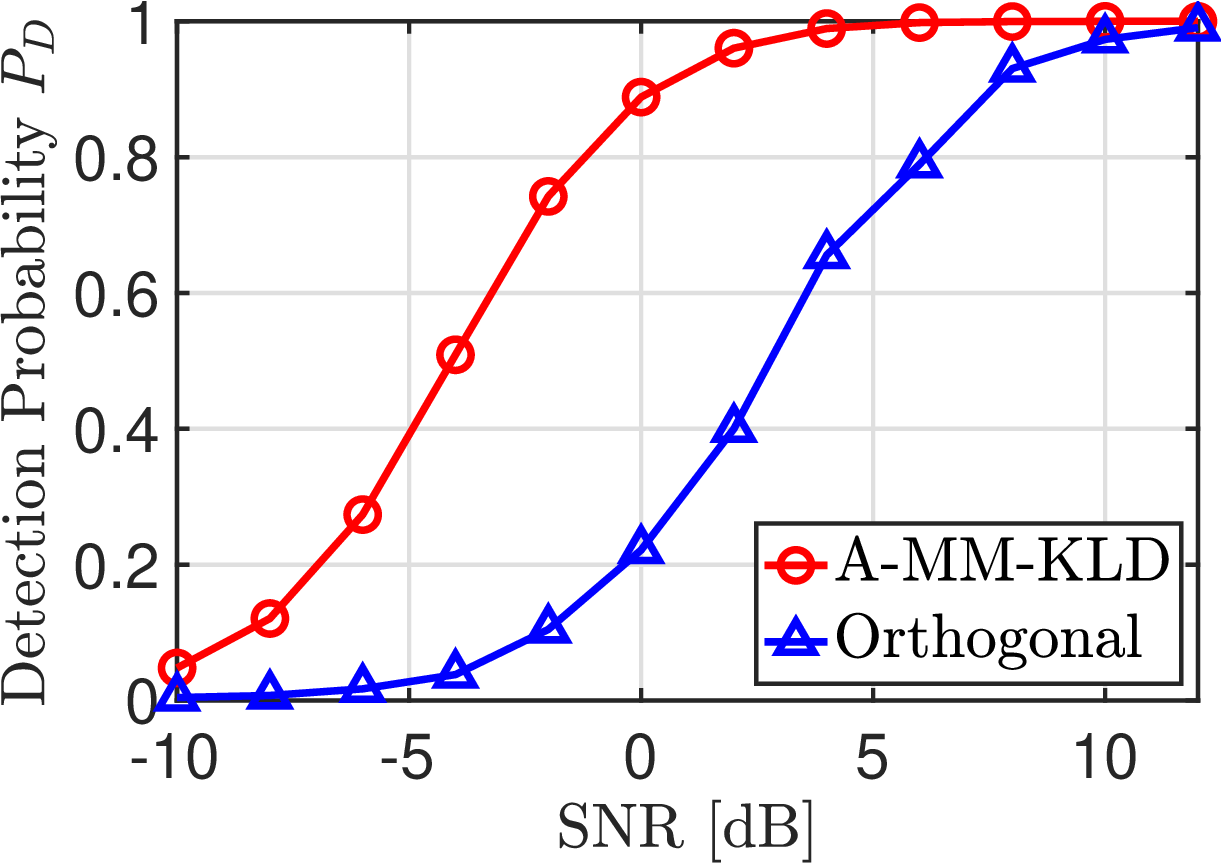}}
    \hfill
    \subfloat[Detection probability versus sequence length ($\mathrm{SNR}=8$~dB).\label{fig:rt_ra_prob_length}]{
        \includegraphics[width=0.46\linewidth]{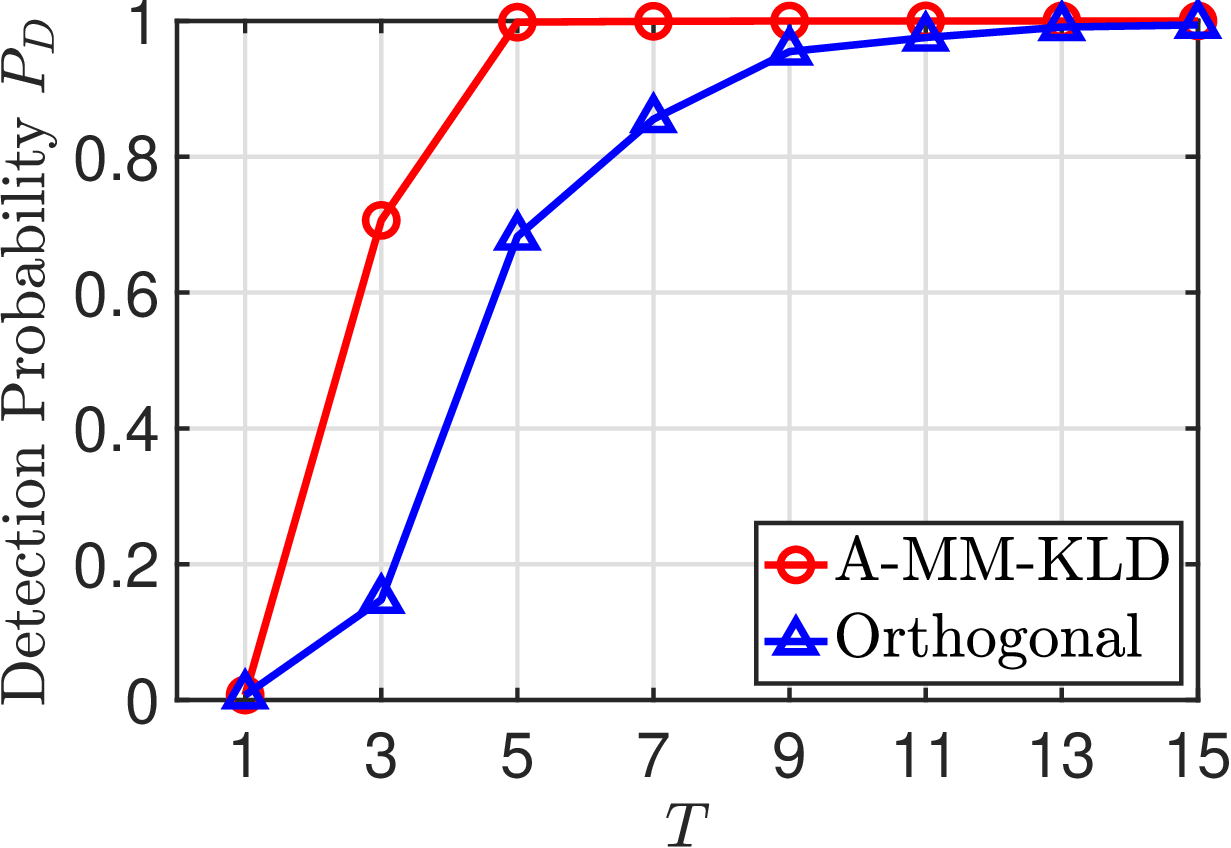}}
    \caption{Geometric-mean detection probability for multiple random access using DeepMIMO.}
    \label{fig:rt_ra_prob}
\end{figure}

We adopt the DeepMIMO ASU Campus scenario at $3.5$~GHz with $N_{t}=N_{r}=4$, $K=4$ users,
and $N_{\mathrm{env}}=8$ distinct site environments, and obtain the per-user covariances
from the corresponding ray-tracing channel samples. The detector, the false-alarm
probability ($\alpha=10^{-3}$), and the geometric-mean detection metric are identical to
those used for the synthetic-covariance experiments above. As shown in Fig.~\ref{fig:rt_ra_prob},
A-MM-KLD again attains a substantial detection-probability gain over the orthogonal
baseline---both as a function of SNR (Fig.~\ref{fig:rt_ra_prob_snr}) and of the sequence length
(Fig.~\ref{fig:rt_ra_prob_length})---reproducing the trends observed under synthetic covariances.
This confirms that the random-access performance gains of the proposed design carry over to
physically grounded, site-specific channel statistics.

\section{Conclusion}
In this paper, we addressed the nonconvex problem of maximizing the KLD for hypothesis testing, a fundamental task in designing optimal sensing systems. We proposed an efficient iterative framework, termed FP-KLD, which leverages a sequence of transforms. This approach systematically reformulates the intractable objective into a sequence of simple concave quadratic subproblems, each of which can be solved efficiently by analyzing its KKT conditions, leading to a Sylvester equation. To further enhance computational efficiency and make the method scalable, we developed an advanced FP-KLD algorithm using nonhomogeneous relaxation. This variant circumvents the high computational cost, resulting in a simple, closed-form update for the sensing waveform at each iteration.

Additionally, we provided a rigorous justification for the monotonic convergence of our algorithms by establishing their equivalence to the MM and BCA frameworks. Furthermore, we demonstrated how the performance of these fixed-point iterations can be substantially improved using standard acceleration techniques. We also illustrated the framework's flexibility by applying it to the ISAC and the multiple random access scenarios. 

The numerical results provided comprehensive insights into the trade-offs among the proposed schemes. 
First, FP-KLD achieved the fastest convergence, empirically validating the tightness of the FP-derived surrogate.
Second, MM-KLD significantly reduced the per-iteration computational burden via nonhomogeneous relaxation, though this came at the cost of an increased number of iterations. 
The A-MM-KLD algorithm successfully combined these advantages; it yielded a remarkable reduction in total runtime compared to state-of-the-art methods while guaranteeing monotonic convergence. 
Our proposed methods also showed strong performance in the joint ISAC waveform design and the random access scenarios.

Several directions exist for extending the proposed framework. One promising way is to apply the A-MM-KLD principle to accelerate MIMO optimizers. In particular, when advanced multiple access schemes (e.g., rate-splitting multiple access) are considered, the resulting objective functions become nonsmooth and highly nonconvex \cite{kim:twc:25, parkGPIRS23, Darktwc25}, often leading to slow convergence of existing algorithms. Adapting our acceleration technique to these scenarios could significantly alleviate the computational burden, paving the way for real-time implementation of complex interference management techniques. 

\bibliographystyle{IEEEtran}
\bibliography{ref_kld_fp}

\end{document}